\documentclass[reqno]{article}

\usepackage{amsthm}
\usepackage{amsmath}
\usepackage{amssymb}
\usepackage{graphicx}
\usepackage{color}
\usepackage{bbm}
\usepackage[top=2.5cm, bottom=2.5cm, left=2.5cm, right=2.5cm]{geometry}
\usepackage{enumerate}
\usepackage{faktor}
\usepackage{nicefrac}
\usepackage[nottoc]{tocbibind}
\usepackage{slashed}
\usepackage{authblk}
\usepackage[normalem]{ulem}
\usepackage{comment}
\usepackage{transparent}

\usepackage[scr=euler]{mathalfa}

\def\mathclap#1{\text{\hbox to 0pt{\hss$\mathsurround=0pt#1$\hss}}}

\newcommand{\loc}{\mathrm{loc}}

\newcommand{\N}{\mathbb{N}}
\newcommand{\R}{\mathbb{R}}
\newcommand{\id}{\mathrm{id}}

\newcommand{\vol}{\mathrm{vol}}

\newcommand{\Sp}{\mathbb{S}}

\newcommand{\uu}{\underline{u}}
\newcommand{\ux}{\underline{x}}
\newcommand{\uchi}{\underline{\chi}}
\newcommand{\ueta}{\underline{\eta}}
\newcommand{\uomega}{\underline{\omega}}

\newcommand{\rd}{\partial}

\begin{document}

\numberwithin{equation}{section}
\newtheorem{theorem}[equation]{Theorem}
\newtheorem{remark}[equation]{Remark}
\newtheorem{assumption}[equation]{Assumption}
\newtheorem{claim}[equation]{Claim}
\newtheorem{lemma}[equation]{Lemma}
\newtheorem{definition}[equation]{Definition}
\newtheorem{corollary}[equation]{Corollary}
\newtheorem{proposition}[equation]{Proposition}
\newtheorem*{theorem*}{Theorem}
\newtheorem{conjecture}[equation]{Conjecture}
\newtheorem{example}[equation]{Example}

\setcounter{tocdepth}{3}

\title{Lipschitz inextendibility of weak null singularities from curvature blow-up}
\author{Jan Sbierski\thanks{School of Mathematics and Maxwell Institute for Mathematical Sciences, 
University of Edinburgh,
James Clerk Maxwell Building,
Peter Guthrie Tait Road, 
Edinburgh, 
EH9 3FD,
United Kingdom; email: Jan.Sbierski@ed.ac.uk}}
\date{\today}

\maketitle

\begin{abstract}
We prove the $C^{0,1}_{\mathrm{loc}}$-inextendibility of weak null singularities without any symmetry assumptions. The proof introduces a new strategy to infer  $C^{0,1}_{\mathrm{loc}}$-inextendibility from the blow-up of curvature. The assumed blow-up is expected to be satisfied for weak null singularities in the interior of generic rotating black holes. Thus, we expect the result presented here to directly contribute to the resolution of the $C^{0,1}_{\mathrm{loc}}$-formulation of the strong cosmic censorship conjecture in a neighbourhood of subextremal Kerr.
\end{abstract}

\tableofcontents

\section{Introduction}

\begin{figure}[h]
 \centering
  \def\svgwidth{11cm}
   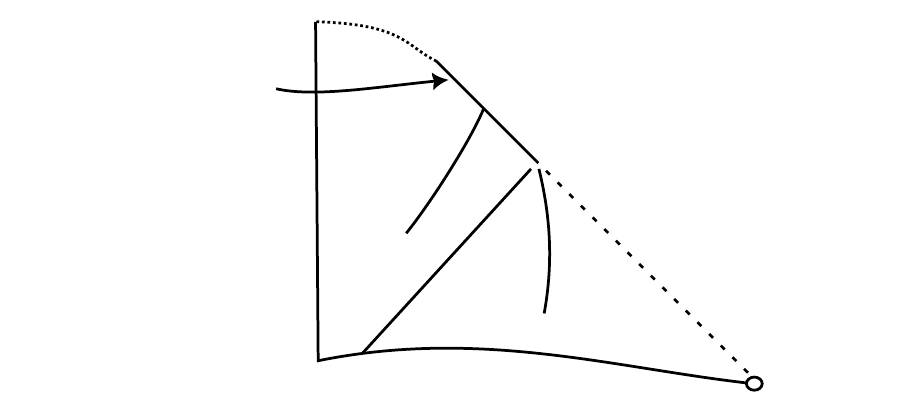 
   \caption{A Penrose-style diagram of a dynamical vacuum black hole settling down to a sub-extremal Kerr black hole in the exterior $I$. Very little is known about the singularity structure in the interior $II$ beyond the weak null singularity. }
   \label{Fig1}
\end{figure}

It is known from the seminal work of Dafermos and Luk \cite{DafLuk17} that dynamical vacuum black holes, which settle down to a subextremal Kerr black hole in the exterior, also have a Cauchy horizon in the black hole interior to which the Lorentzian metric extends continuously, see Figure \ref{Fig1}. Results for the linearised Einstein equations \cite{McNam78}, \cite{Sbie23}, \cite{Gur24} indicate that generically curvature blows up at the Cauchy horizon, turning it into a so-called \emph{weak null singularity}, terminating the classical time-evolution. This scenario is a special case of the much broader strong cosmic censorship conjecture, according to which the maximal globally hyperbolic development of generic asymptotically flat initial data is inextendible as a suitably regular Lorentzian manifold. The main result of this paper, Theorem \ref{MainThm}, shows, that the dynamical black hole interiors of \cite{DafLuk17}, supplemented by a curvature blow-up assumption, are locally Lipschitz inextendible across the weak null singularity. The assumed blow-up of curvature is expected to be generically satisfied, see the discussion in Remark \ref{RemAssump}. Thus, we expect the main theorem to provide one of the ingredients for the resolution of the $C^{0,1}_{\loc}$-formulation of the strong cosmic censorship conjecture in a neighbourhood of subextremal Kerr initial data.

The result presented in this paper is the first low-regularity (i.e., below $C^{1,1}_{\loc}$) inextendibility result for singular spacetimes which are expected to arise from an \emph{open} set of initial data. In particular, no symmetry assumptions on the spacetime are made. Existing low-regularity inextendibility results are either restricted to spacetimes with symmetries \cite{Sbie15}, \cite{GalLin16}, \cite{Sbie22a}, \cite{Sbie23b}, \cite{Ling24}, \cite{Mie24}, with warped product structure \cite{Sbie22a}, or are conditional results  \cite{GalLin16}, \cite{ChrusKli18}, \cite{GraVdBS24}, \cite{Rein24}. An exception are the $C^0$-inextendibility results obtained in \cite{GalLinSbi17}, \cite{MinSuhr19} (see also \cite{GrafLing18}) which are based on geodesic completeness, do not require any symmetries, and do apply for example to the spacetimes arising from small perturbations of Minkowski initial data \cite{ChrisKlainStability}. However, these inextendibility results do not concern singularities.

Of the above cited results on singularities the works \cite{Sbie15}, \cite{GalLin16}, \cite{Sbie23b}, \cite{Ling24}, \cite{Mie24}, \cite{ChrusKli18}, \cite{GraVdBS24} give $C^0$-inextendibility results whereas the work \cite{Sbie22a} gives a $C^{0,1}_{\loc}$-inextendibility result. While for example the works \cite{Sbie15}, \cite{Mie24}, \cite{Sbie23b}, \cite{Ling24} give  methods of proof of $C^0$-inextendibility which in itself do not rely on any symmetry assumptions, \emph{implementing} these methods outside of symmetry is challenging for the reason that it is difficult to gain detailed information on, say, light cones or future one-connectedness in spacetimes without any symmetries. Thus, there is a need for either devising  new simple ways of gaining such information outside of symmetry or for finding new methods of proof whose implementation is also tractable in spacetimes without any symmetries. This paper contributes to the latter. We will come back to this point after we give a rough outline of the idea of the proof.
 
First, however, let us think  about what it means for a spacetime to be $C^{0,1}_{\loc}$-inextendible `across' the weak null singularity, i.e, `across' the Cauchy horizon to which the metric extends continuously by \cite{DafLuk17}: given a $C^{0,1}_{\loc}$-extension, we know by \cite{GalLinSbi17} that there exists at least one inextendible timelike geodesic in the original spacetime which has a limit point in the extension. Hence, $C^{0,1}_{\loc}$-\emph{inextendibility} of the global spacetime can be shown by establishing that none of the inextendible timelike geodesics in the global spacetime  can have a limit point in a $C^{0,1}_{\loc}$-extension. There are many such inextendible timelike geodesics to consider, see Figure \ref{Fig1}: for example $\tau_1$ stays in the black hole exterior, while $\tau_2$ has a limit point on the Cauchy horizon to which the metric extends continuously by \cite{DafLuk17}. This exactly gives us the desired notion: a $C^{0,1}_{\loc}$-extension \emph{across} the weak null singularity is a $C^{0,1}_{\loc}$-extension in which a timelike geodesic $\tau_2$, which has a limit point in the continuous extension of \cite{DafLuk17}, also has a limit point.

We now give the rough outline of the proof, which is by contradiction.
Assume that a $C^{0,1}_{\loc}$-extension across the weak null singularity exists. In the first step we go over from the arbitrary timelike geodesic, which has a limit point in the $C^0$-extension of \cite{DafLuk17} as well as in the assumed $C^{0,1}_{\loc}$-extension, to a well-chosen null geodesic which approaches the same limit points in the respective extensions. This step relies on a result from the precursor \cite{Sbie24a} to this paper.

In the second step we embed this particular null geodesic into a space-filling congruence of null geodesics and use upper bounds on the connection coefficients (in the gauge used in \cite{DafLuk17}) to show that the $C^1$-structure of the $C^{0,1}_{\loc}$-extension has to be locally equivalent to the $C^1$-structure of the $C^0$-extension constructed in \cite{DafLuk17}. Indeed, this local uniqueness problem for $C^{0,1}_{\loc}$-extensions was studied in \cite{Sbie24a} and here, we verify that the assumptions of  \cite{Sbie24a} are met so that we can conclude the local equivalence of $C^1$-structures. An important consequence is that a vector field which extends continuously in the $C^0$-extension of \cite{DafLuk17}, also extends continuously in the assumed $C^{0,1}_{\loc}$-extension. 

In the third step we use the assumed blow-up of curvature: contract curvature with four vector fields which extend continuously in the $C^0$-extension of \cite{DafLuk17} and integrate over a sequence of compact regions asymptotically touching the weak null singularity. We assume that this sequence of integrals diverges. On the other hand, using the coordinates of the $C^{0,1}_{\loc}$-extension, in which the Christoffel symbols remain uniformly bounded, we rewrite these integrals over curvature in terms of these Christoffel symbols. By virtue of the equivalence of the $C^1$-structures, the vector fields in this integral are also regular with respect to those coordinates.\footnote{Indeed, we need that these vector fields are $C^1$, but the equivalence of the $C^1$-structures only gives that the vector fields are continuous. This is the reason for an additional smoothing procedure, which, as a consequence, requires that the integrals over curvature also diverge if we insert vector fields which are $\varepsilon$-close to the original ones.} One can now use Stokes' theorem to convert the integral over first derivatives of Christoffel symboles, which appears in curvature, into a boundary integral over Christoffel symbols. The expression obtained is then manifestly finite and uniformly bounded along the sequence, thus giving the desired contradiction.


The proof given here establishes for the first time an implication that curvature blow-up leads to low-regularity inextendibility.\footnote{For the reverse direction, namely to prove extendibility under curvature bounds, we refer the reader to \cite{Racz10} and references therein. For synthetic curvature blow-up \emph{in} extensions, see \cite{GraKuSa19}.} The subtleties of such a connection were demonstrated in \cite{Ori00}. On the other hand, the $C^{0,1}_{\loc}$-inextendibility of \emph{spherically symmetric} weak null singularities was shown in \cite{Sbie22a} using a method based on direct computation of holonomy. While this method is in principle also applicable to weak null singularities without symmetry assumptions as considered in this paper, it would require, in addition to the control of the $C^1$-structure, (pointwise)  blow-up bounds on one of the connection coefficients (namely $\chi$, see Section \ref{SecRicci}). The new approach presented here is more parsimonious in the sense that only an integrated blow-up bound on curvature is needed, which is easier to establish analytically.

The Lipschitz inextendibility proven in this paper is an important and critical threshold-regularity: firstly, it captures geometrically that the Levi-Civita connection blows up, which is the geometric object at regularity between the metric and curvature. Secondly, Lorentzian causality below Lipschitz regularity changes drastically in comparison to the familiar one of smooth metrics \cite{ChrusGra12}. Nevertheless,  the problem of how to capture geometrically the `rate of the blow up' of the connection, i.e., an inextendibility statement in $C^0 \cap W^{1,p}_{\loc}$, $1<p<\infty$, remains a  challenging open problem (but see Remark \ref{RemRate} for a possible generalisation of the approach presented in this paper).

\subsection*{Acknowledgements}

I acknowledge support through the Royal Society University Research Fellowship URF\textbackslash R1\textbackslash 211216. And I am grateful to an anonymous referee for many helpful suggestions on an earlier version of this manuscript.

\subsection{Preliminaries}

All manifolds considered in this paper are assumed to be smooth.
We begin now by recalling two fundamental definitions and the result which constitutes the starting point for most low-regularity inextendibility results.

\begin{definition} \label{DefExt}
Let $(M,g)$ be a Lorentzian manifold and let $\Gamma$ be a regularity class, for example $\Gamma = C^k$ with $k \in \N \cup \{\infty\}$ or $\Gamma = C^{0,1}_{\loc}$.  A \emph{$\Gamma$-extension of $(M,g)$} consists of a smooth isometric embedding $\iota : M \hookrightarrow \tilde{M}$ of $M$ into a Lorentzian manifold $(\tilde{M}, \tilde{g})$ of the same dimension as $M$ where $\tilde{g}$ is $\Gamma$-regular  and such that $\partial \iota(M) \subseteq \tilde{M}$ is non-empty.

If $(M,g)$ admits a $\Gamma$-extension, then we say that $(M,g)$ is \emph{$\Gamma$-extendible}, otherwise we say $(M,g)$ is \emph{$\Gamma$-inextendbile}.
\end{definition}
Note that for a $\Gamma$-extension of $(M,g)$ to exist the metric $g$ has to be at least $\Gamma$-regular.


\begin{definition} \label{DefFutureBdry}
Let $(M,g)$ be a time-oriented Lorentzian manifold with $g \in C^0$ and let $\iota : M \hookrightarrow \tilde{M}$ be a $C^0$-extension of $M$. The \emph{future boundary of $M$} is the set $\partial^+\iota(M) $ consisting of all points $\tilde{p} \in \tilde{M}$ such that there exists a smooth timelike curve $\tilde{\gamma} : [-1,0] \to \tilde{M}$ such that $\mathrm{Im}(\tilde{\gamma}|_{[-1,0)}) \subseteq \iota(M)$, $\tilde{\gamma}(0) = \tilde{p} \in \partial \iota(M)$, and $\iota^{-1} \circ \tilde{\gamma}|_{[-1,0)}$ is future directed in $M$.
\end{definition}

The next proposition is found in \cite{Sbie18}, Proposition 2.2.

\begin{proposition}\label{PropBoundaryChart}
Let $\iota : M \hookrightarrow \tilde{M}$ be a $C^0$-extension of a time-oriented globally hyperbolic Lorentzian manifold $(M,g)$ with $g \in C^0$ and with Cauchy hypersurface $\Sigma$ --  and let $\tilde{p} \in \partial^+ \iota(M)$. For every $\delta >0$ there exists a chart $\tilde{\varphi} : \tilde{U} \to(-\varepsilon_0, \varepsilon_0) \times  (-\varepsilon_1, \varepsilon_1)^{d}$, $\varepsilon_0, \varepsilon_1 >0$ with the following properties
\begin{enumerate}[i)]
\item $\tilde{p} \in \tilde{U}$ and $\tilde{\varphi}(p) = (0, \ldots, 0)$
\item $|\tilde{g}_{\mu \nu} - \eta_{\mu \nu}| < \delta$, where $\eta_{\mu \nu} = \mathrm{diag}(-1, 1, \ldots , 1)$
\item There exists a Lipschitz continuous function $f : (-\varepsilon_1, \varepsilon_1)^d \to (-\varepsilon_0, \varepsilon_0)$ with the following property: 
\begin{equation}\label{PropF1}
\{(x_0,\underline{x}) \in (-\varepsilon_0, \varepsilon_0) \times (-\varepsilon_1, \varepsilon_1)^{d} \; | \: x_0 < f(\underline{x})\} \subseteq \tilde{\varphi} \big( \iota\big(I^+(\Sigma,M)\big)\cap \tilde{U}\big)
\end{equation} and 
\begin{equation}\label{PropF2}
\{(x_0,\underline{x}) \in (-\varepsilon_0, \varepsilon_0) \times (-\varepsilon_1, \varepsilon_1)^{d}  \; | \: x_0 = f(\underline{x})\} \subseteq \tilde{\varphi}\big(\partial^+\iota(M)\cap \tilde{U}\big) \;.
\end{equation}
Moreover, the set on the left hand side of \eqref{PropF2}, i.e. the graph of $f$, is achronal\footnote{With respect to \emph{smooth} timelike curves.} in $(-\varepsilon_0, \varepsilon_0) \times  (-\varepsilon_1, \varepsilon_1)^{d}$.
\end{enumerate}
\end{proposition}
Given a future boundary point $\tilde{p}$, we call a chart as above a \emph{future boundary chart}.
We also denote by $\tilde{U}_< $ the (image under $\tilde{\varphi}^{-1}$ of) left hand side of \eqref{PropF1}. The region $\{x_0 \leq f(\ux)\}$ including the graph is denoted by $\tilde{U}_\leq$.

\section{The spacetimes considered and gauge choice}

\subsection{The class of spacetimes} \label{SecClassST}

We consider  Lorentzian manifolds $(M,g)$ in a double null gauge: assume $M = (-1,1) \times (-1,0) \times \Sp^2$ with coordinates $(u, \uu)$ on the first two factors and denote with $(\theta^1, \theta^2) = \theta^A$, $A=1,2$, an arbitrary set of smooth coordinates on $\Sp^2$. Let $\gamma(u, \uu)$ be a Riemannian metric on $\Sp^2$ for each $(u, \uu) \in (-1,1) \times (-1,0)$ such that $\gamma_{AB}(u, \uu, \theta^1, \theta^2)$ are at least $C^2$-regular functions of $(u, \uu, \theta^1, \theta^2)$ for each $A,B \in \{1,2\}$. Moreover, let $b(u, \uu)$ be a vector field on $\Sp^2$ for each $(u, \uu) \in (-1,1) \times (-1,0)$ such that $b^A(u, \uu, \theta^1, \theta^2)$ are at least $C^2$-regular functions of $(u, \uu, \theta^1, \theta^2)$ for $A =1,2$. Finally, let $\Omega :(-1,1) \times (-1,0) \times \Sp^2 \to (0, \infty)$ be a positive function that is at least $C^2$-regular. The metric $g$ is  then assumed to be  given in local coordinates $(u, \uu, \theta^A)$  by\footnote{The Einstein summation convention is used throughout the paper.}
\begin{equation} \label{EqGDN}
g = -2 \Omega^2(du \otimes d\uu + d\uu \otimes du) + \gamma_{AB} (d\theta^A - b^A d \uu) \otimes (d\theta^B - b^B d\uu)\;.
\end{equation}
By assumption we have $g \in C^2$. Note that $M$ is orientable -- we fix an orientation and denote the Lorentzian volume form with respect to $g$ by $\mathrm{vol}$.  Moreover, we define a time-orientation on $(M,g)$ by stipulating that $\partial_u$ is future-directed null.

Note that such a double null gauge as in \eqref{EqGDN} exists if, and only if, one can find two solutions $u$ and $\uu$ to the Eikonal equation such that the intersection of their level sets form topological two-spheres.

\subsubsection{Localisation: relation to \cite{DafLuk17}} 

\begin{figure}[h]
 \centering
  \def\svgwidth{4cm}
   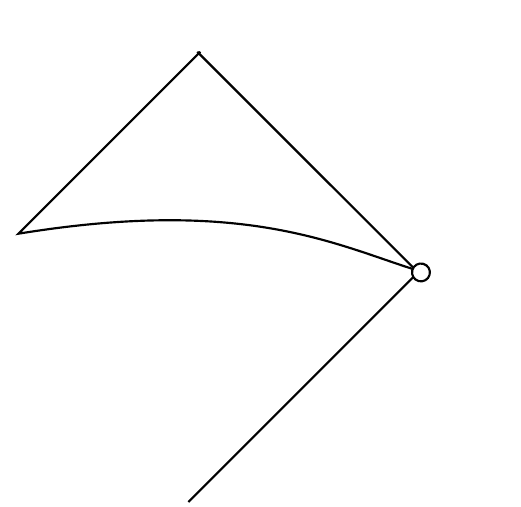 
   \caption{A Penrose-style diagram illustrating the interior of a perturbed sub-extremal Kerr black hole. The shaded region has been constructed in \cite{DafLuk17}.  }
   \label{FigDafLuk}
\end{figure}

The double null gauge used here is the one used in \cite{DafLuk17}. However, in \cite{DafLuk17} a semi-global region of the black hole interior is constructed, see the shaded region in Figure \ref{FigDafLuk}. This shaded region corresponds to the manifold $\mathcal{W} \times \Sp^2$ with $(u, \uu, \theta^A)$ coordinates, where $\mathcal{W} = \{(u, \uu) \in (-\infty, u_f) \times (-\infty, 0) \; | \; u - \frac{1}{2 \kappa} \log (-2 \kappa \uu) \geq C_R\}$, where $u_f \leq -1$, $C_R \in \R$, $\kappa >0$ are constants. The metric in these coordinates takes the form \eqref{EqGDN} and \emph{with respect to the $(u, \uu, \theta^A)$ differentiable structure} extends continuously to $\{\uu = 0\}$, i.e., to $\overline{\mathcal{W}} \times \Sp^2$, where $\overline{\mathcal{W}} = \{(u, \uu) \in (-\infty, u_f) \times (-\infty, 0] \; | \; u - \frac{1}{2 \kappa} \log (-2 \kappa \uu) \geq C_R\}$. The so attached boundary $\rd (\overline{\mathcal{W}} \times \Sp^2)$ at $\{\uu = 0\}$ is  called the Cauchy horizon $\mathcal{CH}^+$. Note that it is a null hypersurface, since $\uu$ is a null coordinate. It is for $\uu \nearrow 0$ that we expect curvature to blow up. Once this is established, we would call $\mathcal{CH}^+$ a \emph{weak null singularity}.

Consider now a timelike geodesic $\tau_2$ with a limit point on $\mathcal{CH}^+$. As discussed in the introduction, in order to show that there are no $C^{0,1}_{\loc}$-extensions across $\mathcal{CH}^+$, we need to show that $\tau_2$ cannot also have a limit point in a $C^{0,1}_{\loc}$-extension. But this is a local problem to study: we may now take a small $(u,\uu)$-chunk of $\mathcal{W} \times \Sp^2$ which contains the future end of $\tau_2$ close to $\mathcal{CH}^+$ and  rescale the $(u, \uu)$-coordinates by a finite constant factor to bring it into the form of $(M,g)$ considered above. Now, since $(M,g)$ isometrically embeds into the semi-global spacetime $\mathcal{W} \times \Sp^2$, any $C^{0,1}_{\loc}$-extension of $\mathcal{W} \times \Sp^2$ in which $\tau_2$ has a future limit point would also give rise to a $C^{0,1}_{\loc}$-extension of $M$ in which $\tau_2$ has a future limit point. So if we show that no such extension exists for $M$, then the same holds for $\mathcal{W} \times \Sp^2$.

This shows that for our purpose it is sufficient  to only consider a finite chunk $(M,g)$ of the interior of the dynamical black hole constructed in \cite{DafLuk17}. Each of those finite chunks $(M,g)$ of the weak null singularity is expected to satisfy the assumptions of Theorem \ref{MainThm}.

\subsubsection{About the concept of a weak null singularity}

Above, we have described an example which we refer to as a \emph{weak null singularity}.
One may attempt to define the notion of a weak null singularity independently of particular examples. One such attempt is as follows: given a (globally hyperbolic) Lorentzian manifold $(M,g)$, a weak null singularity is present if $(M,g)$ admits a $C^0$-extension $\iota : M \hookrightarrow \tilde{M}$ such that part of the boundary of $M$ in $\tilde{M}$ is given by a smooth null hypersurface $\mathcal{N}$. Furthermore, $(M,g)$ is $C^{1,1}_{\loc}$-inextendible across $\mathcal{N}$.

The notion of being $C^{1,1}_{\loc}$-inextendible `across' $\mathcal{N}$ is as explained in the introduction: no $C^{1,1}_{\loc}$-extension exists which provides limit points for timelike geodesics which also have limit points on $\mathcal{N}$ in $\tilde{M}$.

In this case on may choose to identify $\mathcal{N}$ with the weak null singularity. Note that this is in general dependent on the particular choice of $C^0$-extension $\iota : M \hookrightarrow \tilde{M}$, cf.\ \cite{CaSbi24}.

According to this definition, weak null singularities may come in different strengths. Some may be  only $C^{1,1}_{\loc}$-inextendible across $\mathcal{N}$, some may be $C^{0,1}_{\loc}$-inextendible and some may be even more singular. Based on the expected singular behaviour of the metric near the Cauchy horizon $\mathcal{CH}^+$ \emph{in particular coordinates}, we would expect that the weak null singularities in the interior of dynamical black holes are even inextendible in $C^0 \cap W^{1,p}_{\loc}$ for $p >1$.

\subsection{Frame and Ricci coefficients} \label{SecRicci}

We now come back to the class of spacetimes $(M,g)$ introduced in Section \ref{SecClassST} which form the subject of this paper.
The only non-vanishing components of the inverse metric $g^{-1}$ in the $(u, \uu, \theta^A)$ coordinates are 
\begin{equation}
\label{EqInverseM}
(g^{-1})^{AB} = (\gamma^{-1})^{AB}\;, \qquad (g^{-1})^{u \uu} = - \frac{1}{2 \Omega^2} \;, \qquad (g^{-1})^{uA} = - \frac{1}{2\Omega^2} b^A \;.
\end{equation}
Note that $u$ and $\uu$ satisfy $g^{-1}(du, du) = 0 = g^{-1}(d \uu, d\uu)$.

We define two null vectors by 
\begin{equation}
\label{EqDefE3E4}
e_3 := \frac{\rd}{\rd u} \quad \textnormal{ and } \quad e_4 := \frac{1}{\Omega^2} \big( \frac{\rd}{\rd \uu} + b^A \frac{\rd}{\rd \theta^A}\big)\;.
\end{equation} 
Note that we have $g(e_3,e_4) = -2$ and that $e_3 = -2\Omega^2 (d\uu)^\sharp$ and $e_4 =-2 (du)^\sharp$, where $\sharp$ denotes the raising of an index with $g^{-1}$. 

We introduce the following Ricci coefficients with respect to the double null frame $e_3,e_4, e_1,e_2$, where $e_A := \frac{\rd}{\rd \theta^A}$, $A=1,2$:
\begin{align*}
&\chi_{AB} := g(\nabla_{e_A} e_4, e_B)\;, &&\underline{\chi}_{AB}:= g(\nabla_{e_A} e_3, e_B)\;, \\
&\eta_A := -\frac{1}{2} g(\nabla_{e_3} e_A, e_4) \;, && \ueta_A := - \frac{1}{2} g(\nabla_{e_4} e_A, e_3) \;, \\
&  && \uomega := - \frac{1}{4} g(\nabla_{e_3} e_4, e_3) \;. \\
\end{align*}
Note that we have 
\begin{equation} \label{EqOmega}
 4\omega :=- g(\nabla_{e_4} e_3, e_4) =  g(e_3,\nabla_{e_4} e_4) = 0
\end{equation}
and
\begin{equation}\label{EqZeta}
2 \zeta_A := g(\nabla_{e_A} e_4, e_3) = - g(e_4, \nabla_{e_A} e_3) = -g(e_4, \nabla_{e_3} e_A) = 2 \eta_A
\end{equation}
by our choice of double null frame.

Note that $\chi_{AB}$, $\underline{\chi}_{AB}$, $\eta_A$, $\underline{\eta}_A$ can be considered as tensor fields on the two-spheres. The two-spheres have induced Riemannian metric $\gamma$. Thus, it is natural to define the pointwise norms $|\chi|_\gamma = \sqrt{(\gamma^{-1})^{AB} (\gamma^{-1})^{CD} \chi_{AC}\chi_{BD}}$ and $|\eta|_\gamma = \sqrt{ (\gamma^{-1})^{AB} \eta_A \eta_B}$; and similarly for $\uchi$ and $\ueta$. 

We also introduce the vector field $L:= \Omega^2 e_4 = \frac{\rd}{\rd \uu} + b^A \frac{\rd}{\rd \theta^A}$ and denote its integral curve starting at $(u_0, \uu_0, \theta_0) \in M$ by $s \mapsto \sigma_{(u_0, \uu_0, \theta_0)}(s) = \big(u_0, \uu_0 +s, \sigma_{(u_0, \uu_0, \theta_0)}^\theta(s)\big)$. Since $\Sp^2$ is compact, the integral curve exists for $-1<\uu_0 + s <0$. The corresponding flow starting from the hypersurface $\{\uu=-\frac{1}{2}\}$ we denote by 
\begin{equation}
\label{EqFlow}
\Phi : M \cap \{\uu = -\frac{1}{2}\} \times [0, \frac{1}{2}) \to M \cap \{ \uu \geq - \frac{1}{2}\} \,, \qquad \quad \Phi\big((u,  \theta),s\big) = \sigma_{(u, - \frac{1}{2}, \theta)}(s) \;.
\end{equation}
Clearly, $\Phi$ is a diffeomorphism.

Furthermore, we denote with $h$ the auxiliary Riemannian metric on $M$ given by 
\begin{equation}\label{EqDefH}
h := du^2 + d\uu^2 + \gamma(u, \uu)
\end{equation} 
which measures deviation from the $(u, \uu, \theta)$-differentiable structure.
And finally, our convention for the Riemann curvature tensor is
\begin{equation*}
R(X,Y,Z,W) := - g\big( \nabla_X \nabla_Y Z - \nabla_Y \nabla_X Z, W\big) = - g\big( \nabla_X(\nabla_YZ) - \nabla_Y(\nabla_XZ) - \nabla_{[X,Y]} Z, W \big) \;,
\end{equation*}
where $X,Y,Z,W$ are smooth vector fields on $(M,g)$.

\subsection{Causality} \label{SecCausality}

We briefly discuss some elementary causality properties of the class of spacetimes $(M,g)$ introduced in Section 2.1.
First note that $t:= u + 2\uu$ is a time function:  $g^{-1}(dt,dt) <0$ follows directly from \eqref{EqInverseM} and that $t$ increases towards the future follows from $\rd_u(t) >0$.  Its range on $M$ is clearly between $-3$ and $1$.

Given any smooth future directed causal curve $\tau : (0,1) \to M$, we can now parameterise $\tau$ by the time function $t$ to obtain, with slight abuse of notation, $\tau : (t_1, t_2) \to M$ with $-3\leq t_1 < t_2 \leq 1$.
Thus we have $1 = \dot{\tau}^u + 2\dot{\tau}^{\uu}$. Since both terms on the right hand side are positive by the future directedness of $\tau$, they are uniformly bounded. Moreover, we have
\begin{equation}
\label{Eqang}
0 \geq g(\dot{\tau}, \dot{\tau}) = -4 \Omega^2 \dot{\tau}^u \dot{\tau}^{\uu} + \gamma_{AB}(\dot{\tau}^A -b^A \dot{\tau}^{\uu})(\dot{\tau}^B -b^B \dot{\tau}^{\uu})\;.
\end{equation}
We again use the notation $|b|_\gamma = \sqrt{\gamma_{AB} b^A b^B}$ and $|\dot{\tau}|_{\gamma} = \sqrt{\gamma_{AB} \dot{\tau}^A \dot{\tau}^B}$.
Expanding the right hand side of \eqref{Eqang}, dropping a positive term and using Cauchy Schwarz we estimate
\begin{equation*}
\begin{split}
4 \Omega^2 \dot{\tau}^u \dot{\tau}^{\uu} &\geq \gamma_{AB} \dot{\tau}^A \dot{\tau}^B + \gamma_{AB} b^A b^B \dot{\tau}^{\uu} \dot{\tau}^{\uu} - 2 \gamma_{AB} b^A \dot{\tau}^B \dot{\tau}^{\uu} \\
&\geq |\dot{\tau}|_\gamma^2 - 2 \gamma_{AB} b^A \dot{\tau}^B \dot{\tau}^{\uu} \\
& \geq |\dot{\tau}|_\gamma^2 - \frac{1}{2} |\dot{\tau}|_\gamma^2 - 2 (\dot{\tau}^{\uu})^2 | b |_{\gamma}^2\;,
\end{split}
\end{equation*}
which implies the bound on the angular velocities
\begin{equation} \label{EqBoundAngular}
|\dot{\tau}|^2_\gamma \leq 4 (\dot{\tau}^{\uu})^2 | b |_{\gamma}^2 + 8 \Omega^2 \dot{\tau}^u \dot{\tau}^{\uu} \;.
\end{equation}
Assume now in addition that $\tau$ is future inextendible. We claim that this implies $\tau^u(t) \nearrow 1$ or $\tau^{\uu}(t) \nearrow 0$ for $t \nearrow t_2$.\footnote{Both options are allowed, this is not an `either -- or'.}

To prove the claim we recall that $\tau^u$ and $\tau^{\uu}$ are both monotonically increasing; thus, if the claim were not true, we would obtain the limits $\lim_{t \to t_2} \tau^u(t) =: u_0 < 1$ and $\lim_{t \to t_2} \tau^{\uu}(t) =: \uu_0 <0$. But then it follows from \eqref{EqBoundAngular}, from the continuity of $b$, $\gamma$, and $\Omega$, as well as from the uniform boundedness of $\dot{\tau}^u$ and $\dot{\tau}^{\uu}$ that $\dot{\tau}^A$ is uniformly bounded in absolute value for $t \to t_2$. Integration yields that the angular component of $\tau$ also has a limit for $t \to t_2$ and thus $\lim_{t \to t_2} \tau(t) \in M$ exists -- which contradicts the assumption that $\tau$ is future inextendible. This proves the claim.

We can make a similar argument if we assume that $\tau$ is past inextendible. Thus, if we assume that $\tau$ is (past and future) inextendible, then we obtain $\tau^u(t) \nearrow 1$ or $\tau^{\uu}(t) \nearrow 0$ for $t \nearrow t_2$ and $\tau^u(t) \searrow -1$ or $\tau^{\uu}(t) \searrow -1$ for $t \searrow t_1$.

We can now show that $(M,g)$ is globally hyperbolic with Cauchy hypersurface $\Sigma := \{ t = -1\}$: let $\tau$ be an inextendible timelike curve in $M$ which we assume, without loss of generality, to be parameterised by $t$, i.e., $\tau : (t_1, t_2) \to M$ with $-3\leq t_1 < t_2 \leq 1$. By virtue of this parameterisation it is obvious that $\tau$ cannot intersect $\{t = -1\}$ more than once. To show that it \emph{does} intersect $\Sigma$, consider a point, for example $\tau(\frac{t_1 + t_2}{2}) =: (u_0, \uu_0, \theta_0)$, on $\tau$ and first consider the case $u_0 + 2\uu_0 <-1$, i.e., the point lies to the past of $\Sigma$. Since $\tau$ is future inextendible, by the above we have $\tau^u(t) \nearrow 1$ or $\tau^{\uu}(t) \nearrow 0$ for $t \nearrow t_2$. Suppose we have $\tau^u(t) \nearrow 1$. Since $\uu_0 > -1$, we have that $t_2 = \lim_{t \to t_2} [\tau^u(t) + 2 \tau^{\uu}(t) ] \geq 1 + 2 \uu_0 > -1$, which implies that $\tau$ has to cross $\{t = -1\}$. In the case $\tau^{\uu}(t) \nearrow 0$ for $t \nearrow t_2$ we observe that $t_2 = \lim_{t \to t_2}  [\tau^u(t) + 2 \tau^{\uu}(t) ] \geq u_0 >-1$, which again implies that $\tau$ has to cross $\{t=-1\}$.

The case that $u_0 + 2\uu_0 >-1$ is treated analogously. This shows that $\Sigma$ is a Cauchy hypersurface.

\section{The $C^{0,1}_{\loc}$-inextendibility of weak null singularities}

\subsection{Auxiliary results}

\begin{lemma} \label{LemLorentz}
Let $(M,g)$ be a Lorentzian manifold with $g \in C^0$, let $p \in M$ and let $(b_0, b_1, \ldots, b_d)$ and $(f_0, f_1, \ldots, f_d)$ be two orthonormal bases for $T_pM$ with $b_0$ and $f_0$ timelike. Let $L$ denote the Lorentz transformation relating the two bases, i.e., $f_\alpha = L_{\alpha}^{\; \, \beta} b_\beta$. Then $||L||_{\max} := \max_{\alpha \beta} |L_{\alpha}^{\; \, \beta}| \leq |g(b_0, f_0)|$.
\end{lemma}

In other words, the Lorentz transformation relating two orthonormal bases can be uniformly bounded by the inner product of the two timelike basis vectors.

\begin{proof}
The metric $g$ at $p$ expressed in the orthonormal bases takes on the Minkowski form $\eta = \mathrm{diag}(-1,1, \ldots, 1)$. It follows that 
$$ \eta_{\alpha \beta} = g(f_\alpha, f_\beta) = g (L_{\alpha}^{\;\, \gamma} b_\gamma, L_{\beta}^{\;\, \delta} b_\delta) = L_{\alpha}^{\;\, \gamma} L_{\beta}^{\;\, \delta} \eta_{\gamma \delta} \;.$$ 
Let $L^{-1}$ denote the inverse Lorentz transformation $b_\alpha = (L^{-1})_{\alpha}^{\;\, \beta} f_\beta$. Using $(L^{-1})_{\alpha}^{\;\,\beta} L_{\beta}^{\;\, \gamma} = \delta_{\alpha}^{\;\, \gamma}$, we obtain $(L^{-1})_\sigma^{\;\, \alpha} = (\eta^{-1})^{\alpha \beta} L_{\beta}^{\;\, \delta} \eta_{\sigma \delta}$.
Thus we have
\begin{equation}
\label{EqLorentz}
(L^{-1})_{0}^{\;\, 0} = L_{0}^{\;\, 0} \qquad \textnormal{ and } \qquad (L^{-1})_{0}^{\;\, i}  = -L_{i}^{\;\,0} \;,
\end{equation}
where $i = 1, \ldots, d$. Consider
\begin{equation}
\label{EqF0}
f_0 = L_{0}^{\;\,0} b_0 + L_{0}^{\;\,i} b_i
\end{equation} 
and taking the inner product of both sides with $b_0$ gives $L_{0}^{\;\,0} = -g(f_0, b_0)$. Taking the inner product of each side of \eqref{EqF0} with itself gives 
\begin{equation}
\label{EqL00}
 (L_{0}^{\;\,0})^2 = 1 + \sum_i(L_{0}^{\;\,i})^2 \;,
 \end{equation} 
 which provides the desired bound for $L_{0}^{\;\,i}$. To obtain the bound for $L_{i}^{\;\,0}$, we use \eqref{EqLorentz} together with the equivalent of \eqref{EqL00} for $L^{-1}$:
 \begin{equation}
 \label{EqLi0}
 |L_{i}^{\;\,0}|^2 = | (L^{-1})_{0}^{\;\,i}|^2 \leq |(L^{-1})_{0}^{\;\,0}|^2 - 1 = |L_{0}^{\;\,0}|^2 - 1 \;.
 \end{equation}
 Finally, taking the inner product of each side of $f_i = L_{i}^{\;\,0} b_0 + L_{i}^{\;\,j}b_j$ with itself gives $\sum_{j} (L_{i}^{\;\,j})^2 = 1+ (L_{i}^{\;\,0})^2 \leq (L_{0}^{\;\,0})^2$, where we have used \eqref{EqLi0}. This concludes the proof.
\end{proof}

\begin{proposition} \label{PropGeod}
Let $(M,g)$ be a Lorentzian manifold in a double null gauge as in Section \ref{SecClassST} and recall that $g \in C^2$. Assume that the metric \eqref{EqGDN} extends continuously, \emph{with respect to the $(u, \uu, \theta^A)$-coordinates}, as a Lorentzian metric to the manifold with boundary $ M \subseteq \overline{M} := (-1,1) \times (-1,0] \times \Sp^2$.\footnote{This means that $\Omega$ extends as a continuous positive function to $(-1,1) \times (-1,0] \times \Sp^2$, the vector field $b(u, \uu)$ on $\Sp^2$ extends to $(-1,1) \times (-1,0]$ such that $b^A(u, \uu, \theta^1, \theta^2)$ are continuous up to and including $\uu=0$, and $\gamma(u, \uu)$ extends as a Riemannian-metric-on-$\Sp^2$-valued function to $(-1,1) \times (-1,0]$ such that $\gamma_{AB}(u, \uu, \theta^1, \theta^2)$ are continuous up to and including $\uu= 0$.}

Moreover, assume that the following bounds on the Ricci coefficients (with respect to the double null frame from Section \ref{SecRicci})  are satisfied for some $0 < C < 
\infty$:
\begin{equation} \label{EqPropAssumRicci}
\sup\limits_M \{|\uomega| + |\eta|_\gamma + |\ueta|_\gamma + |\uchi|_\gamma\} \leq C \quad \textnormal{ and } \quad \int_{-1}^0 \Big(\sup\limits_{(u, \theta) \in (-1,1)\times \Sp^2} |\chi|_\gamma(u, \uu, \theta) \Big)\, d\uu \leq C \;.
\end{equation}
Let $\tau : [0,t_0) \to M$ be an affinely parameterised future directed timelike geodesic with $\lim_{s \to t_0} \uu\big(\tau(s) \big) = 0$ and $\lim_{s \to t_0} u \big(\tau(s) \big) <1$. Then $\lim_{s \to t_0} \tau(s) \in \overline{M}$ exists, $\tau$ has finite length $0 < t_0 < \infty$, and $h(\dot{\tau}(s), \dot{\tau}(s)) \leq C < \infty$ for all $s \in [0, t_0)$ for some constant $C>0$.
\end{proposition}

We would like to emphasise the convention used in this paper that $C>0$ stands for a generic constant. The exact value of $C$ may change from expression to expression.

This proposition establishes that the affine velocity of timelike geodesics approaching the weak null singularity remains regular with respect to the differentiable structure of $\overline{M}$, i.e., with respect to the differentiable structure of the $C^0$-extension constructed in \cite[Theorem 16.14]{DafLuk17}.

\begin{remark}
Note that $\rd \overline{M} = \overline{M} \cap \{\uu = 0\}$ represents a Cauchy horizon: since $\uu$ is a null coordinate, it is a null hypersurface and its generators, the integral curves of $e_3$, are inextendible causal curves in $\overline{M}$ which do not intersect the Cauchy hypersurface $\{t = -1\}$ of $M \subseteq \overline{M}$.
\end{remark}

\begin{remark}
Note that there indeed exist future directed timelike geodesics in $M$ that approach a point in $\overline{M} \cap \{\uu = 0\}$. To see this, we may extend the boundary $C^0$-extension $\overline{M}$ to a proper $C^0$-extension $\iota : M \hookrightarrow \tilde{M} = (-1,1) \times (-1,1) \times \Sp^2 \supseteq \overline{M}$ in the sense of Definition \ref{DefExt}, see for example the very beginning of the proof of Theorem \ref{MainThm}. The existence of the desired timelike geodesic then follows from \cite[Theorem 2]{GalLinSbi17}.
\end{remark}

\begin{proof}
We first show that $\lim_{s \to t_0} \tau(s) \in \overline{M}$ exists and that the Lorentzian length of $\tau$ is finite. Recall from Section \ref{SecCausality} that we may reparameterise $\tau$ by $t$ to obtain, with slight abuse of notation, $\tau : [t_1, t_2) \to M$ with $-3<t_1 < t_2 \leq 1$. Also recall that in this parameterisation $\dot{\tau}^u$ and $\dot{\tau}^{\uu}$ are positive and uniformly bounded. Exactly the same reasoning as in Section \ref{SecCausality} applies to yield the bound \eqref{EqBoundAngular}. Now, since the metric functions $b$, $\gamma$, $\Omega$ extend continuously to $\{\uu = 0\}$ by assumption, the right hand side of \eqref{EqBoundAngular} is uniformly bounded in $t$. This again implies the uniform boundedness of
 $|\dot{\tau}^A|$. By integration, this firstly gives that $\lim_{s \to t_2} \tau(s) \in \overline{M}$ exists -- and secondly, using again the uniform boundedness of the metric components along $\tau$, it is now straightforward that $L(\tau) = \int_{t_1}^{t_2} \sqrt{ -g(\dot{\tau}(s), \dot{\tau}(s))}\, ds$ is finite. 

We return now to the affine parameterisation $\tau : [0,t_0) \to M$ of $\tau$ and show that its affine velocity vector remains bounded with respect to the $(u, \uu, \theta)$ differentiable structure.  

We complement $e_3$ and $e_4$ with a choice of\footnote{Note that this choice is different to the one made for $e_1$ and $e_2$ in Section \ref{SecRicci}. We will only use this new choice here in this proof.} $e_1$ and $e_2$ along $\tau$ to obtain a normalised double null frame along $\tau$, i.e., $g(e_\mu, e_A) = 0$ for $\mu = 3,4$ and $A=1,2$ and $g(e_A, e_B) = \delta_{AB}$. Expanding $\dot{\tau}$ with respect to this frame we obtain $\dot{\tau} = \dot{\tau}^4 e_4 + \dot{\tau}^3 e_3 + \dot{\tau}^2 e_2 + \dot{\tau}^1 e_1$ with $\dot{\tau}^4 = \Omega^2 \dot{\tau}^{\uu}$ and $\dot{\tau}^3 = \dot{\tau}^u$.
We introduce the future directed timelike vector field $T := e_3 + e_4$ on $M$ and define the energy 
\begin{equation}
\label{EqEnergyGeodesic}
E:= - g(T, \dot{\tau}) = 2 \dot{\tau}^4 + 2\dot{\tau}^3 = 2 \Omega^2 \dot{\tau}^{\uu} + 2 \dot{\tau}^u \;.
\end{equation}
Note that $\dot{\tau}^3$ and $\dot{\tau}^4$ are positive and thus are controlled by $E$. Moreover, the affine parameterisation of $\tau$ implies $-1 = g(\dot{\tau}, \dot{\tau}) = -4 \dot{\tau}^3 \dot{\tau}^4 + (\dot{\tau}^1)^2 +(\dot{\tau}^2)^2 $ and thus
\begin{equation}
\label{EqNormalisedVel}
(\dot{\tau}^1)^2 +(\dot{\tau}^2)^2 \leq 4 \dot{\tau}^3 \dot{\tau}^4 \;.
\end{equation} 
Hence, we obtain
\begin{equation}
\label{EqBoundByEnergy}
 |\dot{\tau}^\alpha| \leq E \qquad \textnormal{ for all } \alpha \in \{1,2,3,4\} \;.
\end{equation}
We compute with the convention that $A,B \in \{1,2\}$ and using $\nabla_{e_4} e_4 = 0$ and \eqref{EqZeta}
\begin{equation} \label{EqRicciGeodesicProof}
\begin{split}
\frac{d}{ds} E &= - g(\nabla_{\dot{\tau}} T, \dot{\tau}) = -\sum_{\alpha, \beta = 1}^4 \dot{\tau}^\alpha \dot{\tau}^\beta \Big[ g(\nabla_{e_\alpha}e_3, e_\beta) + g(\nabla_{e_\alpha} e_4, e_\beta) \Big] \\
&= - \Big(\dot{\tau}^A \dot{\tau}^B \uchi_{AB} + \dot{\tau}^A \dot{\tau}^4 (-2 \zeta_A) + \dot{\tau}^3 \dot{\tau}^4 4 \uomega + \dot{\tau}^4 \dot{\tau}^A 2 \ueta_A \Big) \\
&\quad \, - \Big( \dot{\tau}^A \dot{\tau}^B \chi_{AB} + \dot{\tau}^A \dot{\tau}^3 2 \zeta_A + \dot{\tau}^3 \dot{\tau}^A 2 \eta_A + \dot{\tau}^3 \dot{\tau}^3 (-4 \uomega) \Big) \;.
\end{split}
\end{equation}
With the exception of $\dot{\tau}^A \dot{\tau}^B \chi_{AB}$, every other term on the right hand side can be estimated in absolute value by $C \cdot E^2$, where we use  \eqref{EqBoundByEnergy}, \eqref{EqZeta}, and the first part of \eqref{EqPropAssumRicci}.\footnote{Recall that the symbol $C>0$ does not stand for a particular numerical value, but the constant may change value from line to line.} For the latter note that $\gamma^{-1} = e_1 \otimes e_1 + e_2 \otimes e_2$, so that the components $\underline{\chi}_{AB}$, $\underline{\eta}_A$, etc.\ are controlled by their norms $|\underline{\chi}|_\gamma$, $|\underline{\eta}|_\gamma$, etc.. For the remaining term we use Cauchy Schwarz to obtain $|\dot{\tau}^A\dot{\tau}^B \chi_{AB}| \leq |\dot{\tau}|^2_\gamma |\chi|_\gamma \leq 4 E \dot{\tau}^4 |\chi|_\gamma$, where we have also used \eqref{EqNormalisedVel} and \eqref{EqBoundByEnergy} for the last inequality. Altogether this gives
\begin{equation*}
\frac{d}{ds} E \leq C (E^2 + E \dot{\tau}^4 |\chi|_\gamma) \;.
\end{equation*}
Dividing by $E$ and using the boundedness of $\Omega^2$ along $\tau$ we obtain
\begin{equation*}
\frac{d}{ds} \log E \leq C\Big( E + \dot{\tau}^{\uu} \sup_{(u, \theta) \in (-1,1) \times \Sp^2} |\chi|_\gamma\big(u, \tau^{\uu}(s), \theta\big) \Big) \;,
\end{equation*}
We now integrate over $0 \leq s <t_1 < t_0$. For the first term on the right hand side we use \eqref{EqEnergyGeodesic} and again the uniform boundedness of $\Omega^2$ along $\tau$. These terms may now be integrated explicitly. For the second term on the right hand side we use $\dot{\tau}^{\uu}$ to transform it into an integral over $\uu$ and then use the second part of \eqref{EqPropAssumRicci}. This gives
\begin{equation*}
\log E(t_1) \leq \log  E(0) + C\int\limits_{0}^{t_1}[\dot{\tau}^{\uu}(s) + \dot{\tau}^{u}(s)] \, ds + C \int\limits_{\tau^{\uu}(0)}^{\tau^{\uu}(t_1)} \sup_{(u, \theta) \in (-1,1) \times \Sp^2} |\chi|_\gamma\big(u, \uu, \theta\big) \, d\uu \leq C
\end{equation*}
for all $0<t_1< t_0$. This shows the uniform boundedness of $\dot{\tau}^u(s)$ and $\dot{\tau}^{\uu}(s)$. The uniform boundedness of $\gamma(\dot{\tau}(s), \dot{\tau}(s))$ follows as before from \eqref{EqBoundAngular}. Recalling the definition \eqref{EqDefH} of $h$, this concludes the proof. 
\end{proof}
Note that the important structure of the proof was that the Ricci coefficient in \eqref{EqRicciGeodesicProof}  which is only integrable in $\uu$, i.e., $\chi_{AB}$, comes with a coefficient which is upper bounded in absolute value by $E \cdot \dot{\tau}^4$. 

\subsection{The main theorem}

\begin{theorem} \label{MainThm}
Consider the setting of Proposition \ref{PropGeod} and assume in addition that there is a set of smooth coordinates $(V_{(i)} \subseteq \Sp^2, \theta_{(i)})$ for $\Sp^2$, $i = 1,2$,  with $V_{(1)} \cup V_{(2)} = \Sp^2$ and such that
\begin{equation}
\label{EqAssumpB}
\int\limits_{-1}^0 \sup_{(u, \theta_{(i)}) \in (-1,1) \times V_{(i)}} \big| \frac{\rd}{\rd \theta^B_{(i)}} b^A_{(i)} (u, \uu, \theta_i)\big| \, d \uu  \leq C
\end{equation}
for some $C>0$ and for $i = 1,2$.
Moreover, assume that for every $\overline{p} \in \rd \overline{M}$ and any open neighbourhood $\overline{W} \subseteq \overline{M}$ of $\overline{p}$ there exists $\overline{q} \in \rd \overline{M} \cap \overline{W}$ and a compact neighbourhood $\overline{V} \subseteq \overline{W}$ of $\overline{q}$ such that for all $\uu' \leq 0$ close enough to $0$ the set $\overline{V} \cap \{\uu \leq \uu'\}$ is a compact smooth manifold with corners\footnote{In the sense of \cite{LeeSmooth}.}, and which, moreover, satisfies the following property: there exist continuous vector fields $\overline{X}_i$ on $ \overline{V}$, $i = 1, \ldots, 4$, and an $\varepsilon_0 >0$ such that for any continuous vector fields $\hat{X}_i$ on $\overline{V}$ with $\sup_{\overline{V}} || \hat{X}_i - \overline{X}_i||_h < \varepsilon_0$  we have
\begin{equation}\label{EqThmR}
\Big{|} \int\limits_{\overline{V} \cap \{\uu < \uu_k\}} R(\hat{X}_1, \hat{X}_2, \hat{X}_3, \hat{X}_4) \,\vol \Big{|} \to \infty 
\end{equation}
along a sequence $\uu_k \to 0$.

Then there is no $C^{0,1}_{\loc}$-extension $\hat{\iota} : M \hookrightarrow \hat{M}$ with the property that there is an affinely parameterised, future directed and future inextendible timelike geodesic $\tau : (-t_0,0) \to M$ with $\lim_{s \to 0} \uu\big(\tau(s)\big) = 0$, $\lim_{s \to 0} u\big(\tau(s) \big)<1$, and such that $\lim_{s \to 0} (\hat{\iota} \circ \tau)(s) \in \hat{M}$ exists. Here, $t_0 >0$.
\end{theorem}

See already the left hand side of Figure \ref{FigStep3} on page \pageref{FigStep3} for an illustration of the set-up of Theorem \ref{MainThm}.
Let us remark that the theorem only requires bounds on a subset of the connection coefficients. Bounding a connection strongly relies on the chosen frame -- here we use the frame introduced in Section \ref{SecRicci}. We also briefly comment on the structure of the assumptions in the theorem: assuming the existence of a Lipschitz extension through a point $\overline{p} \in \rd \overline{M}$ in the sense that a timelike geodesic leaves through this point, the first neighbourhood $\overline{W} \subseteq \overline{M}$ arises just as a neighbourhood  of $\overline{p}$ which must also be contained in the extension. And while curvature may be bounded in the direction of isolated points on $\partial \overline{M} \cap \overline{W}$, the next statement is that one can always find an open set of boundary points towards which curvature blows up.

\begin{remark}[Discussion of the assumptions] \label{RemAssump}
All of the assumptions made in the setting of Proposition \ref{PropGeod} are satisfied for each such finite chunk $(M,g)$ of the Cauchy horizon by the work \cite{DafLuk17} of Dafermos and Luk. The bound \eqref{EqAssumpB} is not contained as a formal statement in \cite{DafLuk17} and requires some more work to establish. Note, however, that for weak null singularities only the geometry transversal to the weak null singularity becomes singular -- the tangential geometry remains regular, cf.\ \cite[Theorem 3]{Luk18}.  For the vector fields $\overline{X}_i$ the expectation is that the choice $\overline{X}_1 = \overline{X}_3 = e_4$ and $\overline{X}_2 = e_A$, $\overline{X}_4 = e_B$ gives \eqref{EqThmR}. Indeed, unpublished work\footnote{It may appear as part of forthcoming work.} by the author shows that the result of \cite{Sbie23} can be slightly improved so that the version of \eqref{EqThmR} for the linear Teukolsky field (linearised curvature) is satisfied. Alternatively, this follows from the recent work \cite{Gur24}.
\end{remark}

\begin{remark}[Slight variation of Theorem \ref{MainThm}]\label{RemVariation}
Note that for spacetimes satisfying the vacuum Einstein equations \eqref{EqThmR} reduces to a condition on the Weyl curvature tensor. Thus, in this case, $R(e_4, e_A, e_4, e_B)$  is symmetric in $A$ and $B$ \underline{and} trace-free with respect to $\gamma$ -- it thus has two independent components. We have the following slight variation of Theorem \ref{MainThm} which allows for a combined consideration of these two independent components. The statements are identical up to the following modification:
\begin{quote}
...and which, moreover, satisfies the following property: there exist continuous vector fields $\overline{X}_i^{(j)}$ on $ \overline{V}$, $i = 1, \ldots, 4$, $j=1,2$, and an $\varepsilon_0 >0$ such that for any continuous vector fields $\hat{X}_i^{(j)}$ on $\overline{V}$ with $\sup_{\overline{V}} || \hat{X}_i^{(j)} - \overline{X}_i^{(j)}||_h < \varepsilon_0$  we have
\begin{equation}\label{EqThmR2}
\Big{|} \int\limits_{\overline{V} \cap \{\uu < \uu_k\}} R(\hat{X}_1^{(1)}, \hat{X}_2^{(1)}, \hat{X}_3^{(1)}, \hat{X}_4^{(1)}) + \mathrm{i}R(\hat{X}_1^{(2)}, \hat{X}_2^{(2)}, \hat{X}_3^{(2)}, \hat{X}_4^{(2)}) \,\vol \Big{|} \to \infty 
\end{equation}
along a sequence $\uu_k \to 0$.
\end{quote} 
Here, $\mathrm{i}$ is the complex unit.\footnote{One may have also written \eqref{EqThmR2} as the norm of a vector in $\R^2$.} Setting $\overline{X}_i^{(j)} = e_4$ for $i=1,3$, $j=1,2$, $\overline{X}_2^{(j)} = e_1$ for $j=1,2$, $\overline{X}_4^{(1)} = e_1$, and $\overline{X}_4^{(2)} = e_2$ now combines both components. We expect to apply the theorem in this form in the non-linear setting. The proof of the variation represents only a minor modification of the proof of Theorem \ref{MainThm} and will be addressed below.
\end{remark}

\begin{remark}[Relation to global inextendibility result] \label{RemLocGlob}
The above theorem serves as one piece in a global low-regularity inextendibility statement of generic rotating vacuum black hole spacetimes. Such statements are proven by contradiction. Assuming the existence of a $C^{0,1}_{\loc}$-extension, one knows that there is an  inextendible timelike geodesic in the original spacetime that has a limit point in the extension (Theorem 2 in \cite{GalLinSbi17}, Theorem 3.2 in \cite{Sbie18}, see also \cite{MinSuhr19}). One then leads this statement to a contradiction by  showing that none of the inextendible timelike geodesics can leave into a $C^{0,1}_{\loc}$-extension. Here, one has to take the different geometries along those geodesics into account. Theorem \ref{MainThm} deals with those timelike geodesics approaching the weak null singularity, e.g.\ $\tau_2$ in Figure \ref{Fig1}.
For timelike geodesics lying outside the black hole, for example $\tau_1$ in Figure \ref{Fig1}, one uses their completeness to derive a contradiction \cite{GalLinSbi17}. The geometry in the interior of the black hole to the future of the weak null singularity is still not well-understood, but see for example \cite{BraSmith95}, \cite{VdM23}, \cite{VdM25} for scenarios involving and evidence for additional spacelike singularities.

\end{remark}

\begin{remark}[Structure of condition \eqref{EqThmR}] \label{RemStructureRCond}
The structure of the blow-up condition \eqref{EqThmR} is sharp in the following two senses: one can neither bring the absolute value inside the integral nor can one require \eqref{EqThmR} solely for the particular choice $\overline{X}_i$ of vector fields. Counterexamples to both formulations are presented in Examples \ref{ExCounterAbs} and \ref{ExCounterBar} in Section \ref{SecEx}.
\end{remark}

\begin{remark}[The role of assumption \eqref{EqAssumpB}] \label{RemAs}
The assumption \eqref{EqAssumpB} is used in two ways: on the one hand it is used to control the Ricci-coefficient $g(\nabla_{e_4} e_A, e_B)$. On the other hand it guarantees that the flow $\Phi$ from \eqref{EqFlow} extends as a homeomorphism to $\Phi : M \cap \{\uu = -\frac{1}{2}\} \times [0,\frac{1}{2}] \to \overline{M} \cap \{\uu \geq - \frac{1}{2}\}$. To see the latter we first observe that by the assumption that $g$ extends continuously to $\overline{M}$ we have that the vector field $L$ extends continuously to $\uu = 0$ -- and thus also its integral curves. By a standard continuous dependence argument the flow map $\Phi$  from \eqref{EqFlow} then extends continuously to $M \cap \{ \uu = - \frac{1}{2}\} \times [0, \frac{1}{2}] \to \overline{M} \cap \{\uu \geq - \frac{1}{2}\}$. The surjectivity of this map follows from Peano's existence theorem. For the injectivity we consider two integral curves $\sigma_j(s) = (u, s, \theta_j(s))$, $j = 1,2$, with $\theta_1(-\frac{1}{2}) \neq \theta_2(-\frac{1}{2})$. We want to show that $\lim_{s \to 0} \theta_1(s) \neq \lim_{s \to 0} \theta_2(s)$. For this we can assume that we can choose $-1<s_0<0$ close enough to $0$ such that for $s_0 \leq s <0$  the projections of the two integral curves onto the spheres lie in $ \mathring{V}_{(i)} \subseteq V_{(i)}$ for either $i=1$ or $i=2$, where $\theta_{(i)}^A\big(\mathring{V}_{(i)}) \subseteq \R^2$ is convex. 
We now drop the index from the angular chart for the rest of the argument and compute in coordinates
\begin{equation*}
\frac{d}{ds} || \theta^A_1(s) -  \theta^A_2(s)||_{\R^2} \leq ||b^A\big(\sigma_1(s)\big) - b^A\big(\sigma_2(s) \big)||_{\R^2} \leq \sup_{\substack{\theta \in V_{(i)} \\ B,C \in \{1,2\}}} |\rd_{\theta^B} b^C| \cdot ||\theta_1^A(s) - \theta_2^A(s)||_{\R^2} \;.
\end{equation*}
By \eqref{EqAssumpB} and a Gronwall backwards in $s$ we obtain $$ || \theta^A_1(s_0) -  \theta^A_2(s_0)||_{\R^2} \leq C || \theta^A_1(s) -  \theta^A_2(s)||_{\R^2} $$
for some $C>0$ and for all $s_0 \leq s <0$. This establishes the injectivity.\footnote{For our application to weak null singularities inside generic rotating vacuum black holes this also follows from the estimate above (16.49) in \cite{DafLuk17}.} The continuity of the inverse then follows from the compactness of the domain (for constant $u$). 

The statement of the theorem is adapted to the double null gauge chosen in \cite{DafLuk17}, since presently this is the main application of this result. However, note that a double null gauge as in \cite{Luk18} would slightly change the statement of Theorem \ref{MainThm}; in particular no assumption of the form \eqref{EqAssumpB} would be needed.
\end{remark}

We now give the proof of Theorem \ref{MainThm}. The reader may wish to recall at this point the outline of the idea of the proof given in the introduction.

\begin{proof}
We begin by extending the continuous boundary extension $\overline{M}$ to a continuous extension $\tilde{\iota} : M \hookrightarrow \tilde{M}$ of $(M,g)$ in the sense of Definition \ref{DefExt}: we set $\tilde{M} := (-1,1) \times (-1,1) \times \Sp^2$ and consider the trivial inclusion $\tilde{\iota} : (-1,1) \times (-1,0) \times \Sp^2 \hookrightarrow  (-1,1) \times (-1,1) \times \Sp^2$. By pushing forward $g$ via $\tilde{\iota}$ we obtain a continuous metric on $\overline{M} \subseteq \tilde{M}$ and extend it to $\uu >0$  by setting, with respect to the local $(u, \uu, \theta)$-coordinates,  $\tilde{g}_{\mu \nu} (u, \uu, \theta) := \tilde{g}_{\mu \nu }(u, 0, \theta)$ for $\uu >0$.

To prove the theorem, we now assume that there is a $C^{0,1}_{\loc}$-extension $\hat{\iota} : M \hookrightarrow \hat{M}$ and a future directed and future inextendible timelike geodesic $\tau : (-1,0) \to M$ with $\lim_{s \to 0} \uu\big(\tau(s)\big) = 0$, $\lim_{s \to 0} u\big(\tau(s) \big)<1$, and $\lim_{s \to 0} (\hat{\iota} \circ \tau)(s) =: \hat{p} \in \rd \hat{\iota}(M)$. For the proof it is convenient to go over from the timelike geodesic $\tau$ to a null geodesic (an integral curve of $L$) which also leaves through $\hat{p}$.\footnote{Recall from Remark \ref{RemLocGlob} that Theorem \ref{MainThm} is phrased in terms of a \emph{timelike} geodesic $\tau$ so that it serves as a piece in a \emph{global} $C^{0,1}_{\loc}$-inextendibility statement. In Step 2 below we apply Proposition 5.11 from \cite{Sbie24a} which requires the construction of a space-filling family of causal curves approaching $\{\uu=0\}$ along which parallel transport remains bounded with respect to the differentiable structure of $\overline{M}$ and such that one of them approaches $\tilde{p}$. One could in principle complement $\tau$ by neighbouring timelike geodesics to create such a space-filling family of causal curves. However, this would be more involved than first transitioning to an integral curve of $L$ which also leaves through $\hat{p}$ and then simply making use of the given double null gauge to write down the space-filling family of causal curves in terms of the flow of $L$.} In order to arrange for this, let us first observe that by Proposition \ref{PropGeod} the limit
$\lim_{s \to 0} (\tilde{\iota} \circ \tau)(s) =: \tilde{p} \in \rd\tilde{\iota}(M)$ exists. 
We denote with $\sigma : [-\frac{1}{2}, 0) \to M$ the integral curve of $L$ (null geodesic), starting at parameter value $s= -\frac{1}{2}$ at $(u_0, -\frac{1}{2}, \theta_0) \in M$, such that $\lim_{s \to 0} (\tilde{\iota} \circ \sigma)(s) = \tilde{p}$. There is exactly one such integral curve by Remark \ref{RemAs}, i.e., since $\Phi$ extends as a homeomorphism to $M \cap \{\uu = -\frac{1}{2}\} \times [0,\frac{1}{2}]$.  
\newline

\textbf{Step 1:} \emph{We show that $\lim_{s \to 0} \big(\hat{\iota} \circ \sigma\big)(s) = \hat{p}$.}

We want to apply Proposition 5.1 from \cite{Sbie24a} with $\gamma_1 = \tau$ and $\gamma_2 = \sigma$.\footnote{Also note that the place of $\tilde{M}$ in Proposition 5.1 from \cite{Sbie24a} is taken by $\hat{M}$.}  We start by also parameterising the timelike geodesic $\tau$ by $\uu$ so that, by slight abuse of notation, we obtain
\begin{equation*}
\tau(s) = \big(\tau^u(s), s, \tau^\theta(s) \big) \qquad \textnormal{ and } \qquad \sigma(s) = \big(\tilde{p}^u, s, \sigma^\theta(s) \big) \;.
\end{equation*}
We define the connecting curves $\rho_n$ as follows: first consider the Riemannian metric $\overline{h}:=du^2 + \mathring{\gamma}_2$ on $\R \times \Sp^2$, where $\mathring{\gamma}_2$ denotes the standard round metric on $\Sp^2$. Since $\tau$ and $\sigma$ have the same limit point $\tilde{p}$, we have
\begin{equation*}
\ell_n := d_{\R \times \Sp^2} \Big( \big[ \tau^u(- \nicefrac{1}{n}), \tau^\theta(- \nicefrac{1}{n}) \big], \big[ \tilde{p}^u, \sigma^\theta(- \nicefrac{1}{n}) \big] \Big)  \to 0 \qquad \textnormal{ for } n \to \infty\;.
\end{equation*}
Let now $\overline{\rho}_n : [0, \ell_n] \to \R \times \Sp^2$ be the shortest curve in $(\R \times \Sp^2, \overline{h})$ connecting $\big[ \tau^u(- \nicefrac{1}{n}), \tau^\theta(- \nicefrac{1}{n}) \big]$ with $\big[ \tilde{p}^u, \sigma^\theta(- \nicefrac{1}{n}) \big]$, parameterised by arclength. For $n$ large enough, there is a unique such  curve and thus $\overline{\rho}_n$ is well-defined. We then define $\rho_n : [0, \ell_n] \to M$ by
\begin{equation} \label{EqDefRho}
\rho_n(s) = \begin{cases} \big( \overline{\rho}_n^u(s), - \frac{1}{n} -s , \overline{\rho}_n^\theta(s) \big) \quad &\textnormal{ for } 0 \leq s \leq \frac{1}{2} \ell_n \\
\big( \overline{\rho}_n^u(s), - \frac{1}{n} -\ell_n + s , \overline{\rho}_n^\theta(s) \big) \quad &\textnormal{ for } \frac{1}{2} \ell_n \leq s \leq \ell_n \;.
\end{cases}
\end{equation}
Note that $\rho_n$ connects $\tau(-\nicefrac{1}{n})$ with $\sigma(-\nicefrac{1}{n})$.
As will become clear later, it is important that the connecting curves $\rho_n$ move also a little backwards and forwards in $\uu$ -- otherwise their gap-length would not be finite in general. For the set-up see also Figure \ref{FigStep1}.
\begin{figure}[h] 
  \centering
  \def\svgwidth{5cm}
   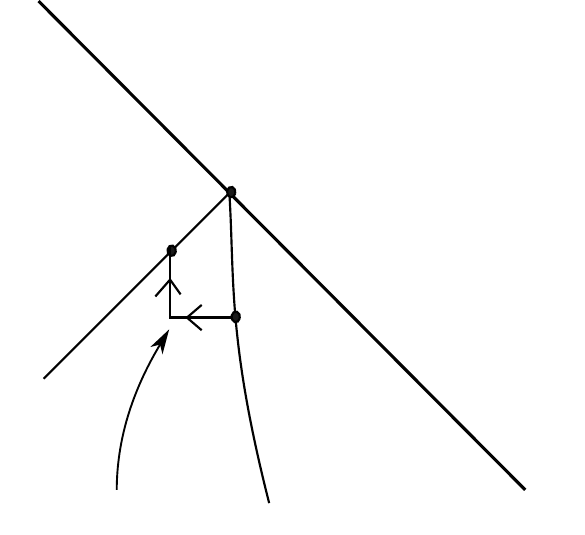 
      \caption{The curves and frames in $\tilde{M}$ used in Step 1.} \label{FigStep1}
\end{figure}

We now introduce a reference orthonormal frame $f_\alpha$ given by $f_0 := \frac{1}{2}(e_3 + e_4)$, $f_1:=\frac{1}{2}(e_3 - e_4)$, and a smooth choice of $f_2$ and $f_3$.  The exact choice of $f_2$ and $f_3$ is not important and we only need this frame on the union of the curves $\tau$, $\sigma$, and $\rho_n$, so that $f_2$ and $f_3$ can be smoothly defined.\footnote{For example $f_2$ and $f_3$ can be constructed as follows: consider the projection of the paths $\tau$, $\sigma$, and $\rho_n$ to the $\Sp^2$ in $M = (-1,1) \times (-1,0) \times \Sp^2$. By Sard's theorem the image of the projection of each of those paths has Lebesgue measure zero in $\Sp^2$. Thus, the countable union of the images in $\Sp^2$ of those paths is still a null set. Hence, there exists a point $\omega_0 \in \Sp^2$ which does not lie in the projection onto $\Sp^2$ of any of those paths. But on $(-1,1) \times (-1,0) \times (\Sp^2 \setminus \{\omega_0\})$ there exists a global orthonormal frame.}

Since $\overline{h}\big(\dot{\overline{\rho}}_n(s), \dot{\overline{\rho}}_n(s)\big) = 1$, it in particular follows that there is a constant $C>0$, uniform in $n \in \N$ and $s \in [0, \ell_n]$, such that 
\begin{equation}
\label{EqUniformBoundRho}
|g\big(\dot{\rho}_n(s), f_\alpha\big) | \leq C\;.
\end{equation}
In particular, if we expand $\dot{\rho}_n(s) = \dot{\rho}_n^\alpha(s) f_\alpha|_{\rho_n(s)}$, then the components $\dot{\rho}_n^\alpha$ with respect to the $f_\alpha$ frame are uniformly bounded.

Before we continue, we refer the reader unfamiliar with the concept of `generalised affine parameter length' (in short: gap-length) of a curve to Section 8.1 in \cite{HawkEllis} or the beginning of Section 4 in \cite{Sbie24a}. Note in particular that for geodesics an affine parameter is also a generalised affine parameter. 

One of the assumptions of Proposition 5.1 in \cite{Sbie24a} is that the gap-length of $\sigma$ is finite. Since $\nabla_{e_4} e_4 = 0$ and $\dot{\sigma} = L = \Omega^2 e_4$, $\sigma$ is a reparameterised null geodesic. It thus suffices to show that its affine parameter length is finite. By \eqref{EqDefE3E4}, the affine parameter $s$ of $\sigma$ associated to $e_4$ is determined by $\frac{d \uu}{ds} = \frac{1}{\Omega^2}$ -- and thus the uniform boundedness of $\Omega^2$ along $\sigma$ implies the finiteness of the affine parameter length.

Next, we introduce a parallely propagated frame $d_\alpha$ along $\sigma$ and show that it is related to the $f_\alpha$ frame by uniformly bounded Lorentz transformations. We fix $d_\alpha$ by demanding $d_\alpha \big(\sigma(-\frac{1}{2}) \big) = f_\alpha\big(\sigma(- \frac{1}{2}) \big)$. We claim that we have
\begin{equation}
\label{EqClaimDF}
|g(d_0, f_0)| \leq C
\end{equation}
for some $C>0$ uniform along $\sigma$. By Lemma \ref{LemLorentz} this implies that the two frames are related by uniformly bounded Lorentz transformations.

To prove \eqref{EqClaimDF}, we write $d_0 = d_0^{\;\,0} f_0 + d_0^{\;\,i} f_i$, where $i$ ranges from $1$ to $3$. Since parallel transport preserves the spacetime inner product, we have 
\begin{equation}
\label{EqBoundDi}
-1 = - (d_0^{\;\,0})^2 + \sum_{i=1}^3 (d_0^{\;\,i})^2 \;.
\end{equation}
Since we also have $d_0^{\;\, 0}(-\frac{1}{2}) =1$, this gives  $d_0^{\;\,0}(s) \geq 1$ for all $s \in [-\frac{1}{2}, 0)$. By definition we have $\nabla_{\dot{\sigma}}d_0 = 0$ and thus we get
$$0 = g(\nabla_{\dot{\sigma}} d_0, f_0) = - \frac{d}{ds} d_0^{\;\,0} + d_0^{\;\,i} \cdot g(\nabla_{\dot{\sigma}} f_i, f_0) \;,$$
where we have used $g(f_i, f_0) = 0$ and $g(\nabla_{\dot{\sigma}} f_0, f_0) = 0$, which follows from $g(f_0, f_0) = -1$.
Using \eqref{EqBoundDi} to bound $d_0^{\;\,i}$ in terms of $d_0^{\;\,0}$ and dividing by $d_0^{\;\,0}$ we get (see also justification below) $$\frac{d}{ds} \log d_0^{\;\,0} \leq \sum_{i=1}^3 | g( \nabla_{\dot{\sigma}} f_i, f_0)| \leq C \sum_{i=1}^3 | g( \nabla_{e_4} f_i, f_0)| \leq C \sum_{i=1}^3 | g(  f_i,\nabla_{e_4} e_3)| \leq C \sum_{i=2,3} |\underline{\eta}(f_i)|  \leq C |\underline{\eta}|_\gamma\;,$$
where we have also used $\dot{\sigma} = L = \Omega^2 e_4$ and that $\Omega^2$ is bounded along $\sigma$ in the second inequality, $g(f_i, f_0) = 0$ together with the definition of $f_0$ and $\nabla_{e_4} e_4 = 0$ in the third inequality. For the fourth inequality we used again $\nabla_{e_4} e_4 = 0$ and $g(e_3, e_3) = 0$ to see that the term with $i=1$  vanishes and the definition of $\underline{\eta}$. Since $f_2$ and $f_3$ lie in the orthogonal complement of $e_3$ and $e_4$ and are thus tangential to the $\Sp^2$, we have $\gamma^{-1} = f_2 \otimes f_2 + f_3 \otimes f_3$, which proves the last step.
By \eqref{EqPropAssumRicci}, this now proves the claim \eqref{EqClaimDF}.

Next we consider a parallely propagated orthonormal frame $b_\beta$ along $\tau$ such that $b_0 =\dot{\tau}$. By Proposition \ref{PropGeod} we have that $h(\dot{\tau}(s), \dot{\tau}(s))$ is uniformly bounded, which, together with the continuity of the metric up to $\{\uu = 0\}$ -- and thus the uniform boundedness of its components along $\tau$ -- gives
\begin{equation}
\label{EqBF}
|g(b_0, f_0)| \leq C
\end{equation}
for some constant $C>0$ which is uniform along $\tau$.

We now claim that $L_{\mathrm{gap}, b_\beta} (\rho_n) \to 0$ for $n \to \infty$, which is another assumption in Proposition 5.1 in \cite{Sbie24a}. To prove this, we first establish that for the parallel transport of $b_\beta$ along $\rho_n$ we have $|g(b_0, f_0)| \leq C$ uniformly in $s \in [0, \ell_n]$ and $n \in \N$. 

We proceed as before. Writing $b_0 = b_0^{\;\,0} f_0 + b_0^{\;\,i} f_i$ it follows from $ 0 = g(\nabla_{\dot{\rho}_n} b_0, f_0)$ in the same way as before that $\frac{d}{ds} \log b_0^{\;\,0} \leq \sum_i | g(\nabla_{\dot{\rho}_n} f_i, f_0) |$. Using \eqref{EqUniformBoundRho} this gives 
\begin{equation}
\label{EqLogB}
\frac{d}{ds} \log b_0^{\;\,0} \leq C\sum_{i=1}^3 \sum_{\alpha = 0}^3 | g(\nabla_{f_\alpha} f_i, f_0) | = C\sum_{i=1}^3 \sum_{\alpha = 0}^3 | g( f_i, \nabla_{f_\alpha} f_0) | \;.
\end{equation}
Using the definition of the $f_\alpha$ it follows that in order to bound the right hand side of \eqref{EqLogB} it suffices to bound $|g(e_\beta, \nabla_{e_\alpha} e_3)|$ and $|g(e_\beta, \nabla_{e_\alpha} e_4)|$ for $\alpha, \beta \in \{1,2,3,4\}$.\footnote{The reader may wish to consider here $e_1 = f_1$ and $e_2 = f_2$ to simplify the reasoning, but this is not essential.} Let's consider those cases:
\begin{enumerate}
\item $|g(e_\beta, \nabla_{e_\alpha} e_3)|$
\begin{itemize}
\item $\beta = 1,2$: for $\alpha = 1,2$ this is controlled by $|\underline{\chi}|_\gamma$. For $\alpha = 3$ it vanishes and for $\alpha = 4$ it is controlled by $|\underline{\eta}|_\gamma$.
\item $\beta = 3$: those terms vanish.
\item $\beta = 4$: for $\alpha = 1,2$ we use \eqref{EqZeta} to see that it is controlled by $|\eta|_\gamma$. For $\alpha = 3$ it is controlled by $|\underline{\omega}|$, and for $\alpha = 4$ it vanishes.
\end{itemize}
\item $|g(e_\beta, \nabla_{e_\alpha} e_4)|$
\begin{itemize}
\item $\beta = 1,2$: for $\alpha =1,2$ this is controlled by $|\chi|_\gamma$. For $\alpha = 3$ this is controlled by $|\eta|_\gamma$ and for $\alpha = 4$ this vanishes.
\item $\beta = 3$: for $\alpha = 1,2$ we use \eqref{EqZeta} to see that it is controlled by $|\eta|_\gamma$. For $\alpha = 3$ it is controlled by $|\underline{\omega}|$ and for $\alpha = 4$ it vanishes.
\item $\beta = 4$: those terms vanish.
\end{itemize}
\end{enumerate}
We now use the bounds \eqref{EqPropAssumRicci} and integrate \eqref{EqLogB} along $\rho_n$: with the exception of $|\chi|_\gamma$, all other terms from the above list are thus bounded by $C \cdot \ell_n$. For $|\chi|_\gamma$ we recall the definition \eqref{EqDefRho} of $\rho_n$ and estimate by taking the supremum in $u$ and $\theta$ to obtain
$$\Big| \log (b_0^{\;\,0})|_{\rho_n(s)} - \log(b_0^{\;\,0})|_{\tau(- \nicefrac{1}{n})} \Big| \leq C\Big(\ell_n + 2 \int\limits_{-\nicefrac{1}{n} - \nicefrac{\ell_n}{2}}^{- \nicefrac{1}{n}} \sup_{u, \theta}|\chi|_\gamma(\uu) \, d\uu \Big) \to 0$$
for all $s \in [0, \ell_n]$ and $n \to \infty$. Since $0<b_0^{\;\,0}|_{\tau(- \nicefrac{1}{n})} \leq C$ uniformly in $n \in \N$ by \eqref{EqBF}, this gives 
\begin{equation}
\label{EqBRho}
-g(b_0, f_0)\big|_{\rho_n(s)} = b_0^{\;\,0}|_{\rho_n(s)} \leq C
\end{equation}
uniformly in $n$ and $s$. Recall here our convention that $C>0$ stands for a generic constant whose value may change from line to line.

By Lemma \ref{LemLorentz} the Lorentz transformations relating $b_\beta $ and $f_\alpha$ along $\rho_n$ are uniformly bounded. Since by \eqref{EqUniformBoundRho} the components of $\dot{\rho}_n = \dot{\rho}_n^\alpha f_\alpha$ with respect to the frame $f_\alpha$ are uniformly bounded, we also obtain that the components of $\dot{\rho}_n = \dot{\rho}_n^\beta b_\beta$ with respect to the frame $b_\beta$ are uniformly bounded in $n$ and $s$. But this gives $L_{\mathrm{gap}, b_\beta}(\rho_n) = \int_0^{\ell_n} \sqrt{ \sum_\beta |\dot{\rho}_n^\beta(s) |^2} \, ds \to 0$ for $n \to \infty$ since $\ell_n \to 0$.

Finally, note that by Lemma \ref{LemLorentz} and \eqref{EqClaimDF} together with \eqref{EqBRho} the Lorentz transformations relating the orthonormal bases $b_\beta|_{\rho_n(\ell_n)}$ with $f_\alpha|_{\rho_n(\ell_n)}$ are uniformly bounded along $\rho_n(\ell_n) = \sigma(-\nicefrac{1}{n})$ as well as the Lorentz transformations relating $f_\alpha$ with $d_\alpha$ along $\sigma$. Thus the Lorentz transformations relating $b_\beta$ with $d_\alpha$ on $\rho_n(\ell_n) = \sigma(-\nicefrac{1}{n})$ are uniformly bounded in $n$. This verifies the assumptions of Proposition 5.1 in \cite{Sbie24a} and we infer $\lim_{s \to 0} ( \iota \circ \sigma)(s) = \hat{p}$.
\newline

\textbf{Step 2:} \emph{Let $\id := \hat{\iota} \circ \tilde{\iota}^{-1} : \tilde{\iota}(M) \to \hat{\iota}(M)$ denote the identification map. We show that there exists a future boundary chart $\hat{\varphi} : \hat{U} \to (-\varepsilon_2, \varepsilon_2) \times (-\varepsilon_3, \varepsilon_3)^3$ around $\hat{p} \subseteq \hat{M}$ and neighbourhoods $\tilde{W} \subseteq \tilde{M}$ of $\tilde{p}$ and $\hat{W} \subseteq \hat{U}$ of $\hat{p}$ such that $\id|_{\tilde{W}_<} : \tilde{W}_< \to \hat{W}_<$ is a diffeomorphism and extends as a $C^1$-regular isometric diffeomorphism to $\id|_{\tilde{W}_\leq} : \tilde{W}_\leq \to \hat{W}_\leq$.
Here we have set\footnote{See the definition of $\hat{U}_<$ and $\hat{U}_\leq$ below Proposition \ref{PropBoundaryChart}.} $\tilde{W}_< := \tilde{W} \cap \{ \uu <0\}$, $\tilde{W}_\leq := \tilde{W} \cap \{\uu \leq 0\}$ and $\hat{W}_< := \hat{W} \cap \hat{U}_<$, $\hat{W}_\leq := \hat{W} \cap \hat{U}_\leq$. }

We show the claim by verifying the assumptions of Proposition 5.11 in \cite{Sbie24a}. The setting of Proposition 5.11 is in terms of a future boundary chart $\tilde{\varphi} : \tilde{U} \to (-\varepsilon_0, \varepsilon_0) \times (-\varepsilon_1, \varepsilon_1)^3$ around $\tilde{p} \in \tilde{M}$. Such a chart can be easily constructed  by  first making an affine coordinate transformation $y^\alpha = B^\alpha_{\;\, \mu} x^\mu + p^\alpha$ from the $y^\alpha = (u, \uu, \theta^A)$ coordinates to new $x^\mu$ coordinates: firstly, the real $4\times 4$ matrix $B^\alpha_{\;\, \mu}$ can be chosen such that $\tilde{g}_{\mu \nu}|_{\tilde{p}} = B^\alpha_{\;\, \mu} B^\beta_{\;\, \nu} \tilde{g}_{\alpha \beta}|_{\tilde{p}} \overset{!}{=} \eta_{\mu \nu}$, where $\eta_{\mu \nu} = \mathrm{diag}(-1,1,1,1)$, and such that $\frac{\rd}{\rd x^0}|_{\tilde{p}}$ is future directed. The condition $i)$ in Proposition \ref{PropBoundaryChart} can then be arranged by choice of the vector $p^\alpha$ in $\R^4$. Using the continuity of the metric $\tilde{g}$, condition $ii)$ follows from $\tilde{g}_{\mu \nu}|_{\tilde{p}} = \eta_{\mu \nu}$ after making $\varepsilon_0, \varepsilon_1>0$ sufficiently small. Finally, condition $iii)$ follows from writing $\{\uu = 0\}$ as $0 = B^{\uu}_{\;\, \mu} x^\mu - p^{\uu}$ together with $B^{\uu}_{\;\, 0} \neq 0$, which in turn follows from $\frac{\rd}{\rd x^0} = B^\alpha_{\; \, 0} \frac{\rd}{\rd y^\alpha}$ and $d\uu (\frac{\rd}{\rd x^0}) \neq 0$ by virtue of $\uu$ being a null coordinate and $\frac{\rd}{\rd x^0}$ timelike.   Also note that we have $\tilde{U}_< = \tilde{U} \cap \{\uu <0\}$.

Recall the definition \eqref{EqFlow} of $\Phi$ and that  $\Phi\big((u_0, \theta_0), \frac{1}{2}\big) = \tilde{p}$. We now take a coordinate chart $(V_{(i)}, \theta_{(i)})$ in which $\theta_0$ is contained and we drop the index of this chart. Consider a coordinate ball $B_{\rho}\big((u_0, \theta_0)\big) \subseteq \R^3$ around $(u_0, \theta_0)$ with respect to $(u, \theta^A)$ coordinates and then make $1 - |u_0| >\rho>0$ so small  and choose $0 \leq s_0 < \frac{1}{2}$ close enough to $\frac{1}{2}$ such that $\Phi \Big( B_\rho \big((u_0, \theta_0)\big), [s_0, \frac{1}{2}] \Big) \subseteq \tilde{U}$. This uses the fact that $\Phi$ extends continuously to $M \cap \{\uu =- \frac{1}{2}\} \times [0, \frac{1}{2}]$. We thus have that $\Phi : B_\rho \big((u_0, \theta_0)\big) \times [s_0, \frac{1}{2}) \to \tilde{U}_<$ is a diffeomorphism onto its image, extends as a homeomorphism  $\Phi : B_\rho \big((u_0, \theta_0)\big) \times [s_0, \frac{1}{2}] \to \tilde{U}_\leq$ onto its image, and that $\partial_s x^\mu\big(\Phi(u, \theta, s)\big) = L^\mu$ is uniformly bounded.

After making $\rho>0$ smaller and choosing $0\leq s_0 < \frac{1}{2}$ closer to $\frac{1}{2}$ if necessary, we can assume that the closure of the projection of $\Phi\Big(B_\rho \big((u_0, \theta_0)\big) \times [s_0, \frac{1}{2})\Big)$ onto $\Sp^2$ is contained in one of the charts $(V_{(i)}, \theta_{(i)})$, of which we again drop the index in the following. We complement $f_0$ and $f_1$  to get a smooth orthonormal frame $f_\alpha$ on $\Phi\Big(B_\rho \big((u_0, \theta_0)\big) \times \{s_0\}\Big)$ and we define an orthonormal frame $d_\alpha$ on $\Phi\Big(B_\rho \big((u_0, \theta_0)\big) \times [s_0, \frac{1}{2})\Big)$ by parallely propagating $d_\alpha$ along $L$ with initial condition $d_\alpha = f_\alpha$.

\underline{Claim:} The orthonormal frame $d_\alpha$ extends continuously to $\Phi\Big(B_\rho \big((u_0, \theta_0)\big) \times [s_0, \frac{1}{2}]\Big)$. 

To prove the claim, we use the smooth frame field $e_\alpha$ from Section \ref{SecRicci} on the domain under consideration, where $e_A = \rd_{\theta^A}$, for $A=1,2$. The dual frame is given by
\begin{equation*}
\omega^A = g\big((d\theta^A)^{\sharp_\gamma}, \cdot\big) \;, \quad \omega^3 = - \frac{1}{2} g(e_4, \cdot)\;, \quad \omega^4 = - \frac{1}{2} g(e_3, \cdot) \;,
\end{equation*}
where $\sharp_\gamma$ denotes the raising of an index with respect to the metric $\gamma$ on the spheres. We then expand $d_\beta = d_{\beta}^{\;\,\alpha} e_\alpha$ and contracting $0 = \nabla_L d_\beta = \frac{d}{ds} d^{\;\,\alpha}_{\beta} \cdot e_\alpha + d^{\;\,\alpha}_{\beta} \cdot \nabla_Le_\alpha$ with $\omega^\rho$ gives
\begin{equation*}
\frac{d}{ds} d^{\;\,\rho}_{\beta} = - d^{\;\,\alpha}_{\beta} \omega^\rho\big(\nabla_L e_\alpha\big) =: -  d^{\;\,\alpha}_{ \beta} \mathbb{A}^{\;\,\rho}_{ \alpha}
\end{equation*}
with 
\begin{equation*}
\begin{split}
&\mathbb{A}^{\;\,4}_{ \alpha} = - \frac{1}{2} g(e_3, \nabla_L e_\alpha) = \Omega^2 \frac{1}{2} g(\nabla_{e_4} e_3 , e_\alpha) = \begin{cases} \Omega^2 \ueta_A \quad &\textnormal{ for } \alpha = A \\ 0 \quad &\textnormal{ for } \alpha \in \{3,4\}  \end{cases} \\
& \mathbb{A}^{\;\,3}_{ \alpha} = - \frac{1}{2} g(e_4, \nabla_L e_\alpha) = 0 \quad \textnormal{ for } \alpha \in \{1,2,3,4\} \\
&\mathbb{A}^{\;\,A}_{ \alpha} = \gamma^{AB} g(e_B, \nabla_L e_\alpha) = \begin{cases} \gamma^{AB} g(e_B, \nabla_{e_C} L) + \gamma^{AB} g(e_B, [L, e_C]) = \Omega^2 \gamma^{AB}  \chi_{BC} - \rd_{\theta^C} b^A \quad &\textnormal{ for } \alpha = C \\
\Omega^2 \gamma^{AB} g(e_B, \nabla_{e_4} e_3) = 2 \Omega^2 \gamma^{AB} \ueta_B &\textnormal{ for } \alpha = 3 \\
0 &\textnormal{ for } \alpha = 4 \;. \end{cases}
\end{split}
\end{equation*}
Assumptions \eqref{EqPropAssumRicci} and \eqref{EqAssumpB}, the continuity of $\Omega^2$ together with $1- |u_0| > \rho >0$, as well as  the continuity of $\gamma_{AB}$ together with the closure of the projection of $\Phi\Big(B_\rho \big((u_0, \theta_0)\big) \times [s_0, \frac{1}{2})\Big)$ onto $\Sp^2$ being contained in one angular chart, give
\begin{equation}
\label{EqIntA}
\int_{-\frac{1}{2}}^0 \sup\limits_{(u, \theta) \in B_\rho\big((u_0, \theta_0) \big)} ||\mathbb{A}^{\;\,\rho}_{ \alpha}|_{\Phi((u,\theta),s)}|| \, ds \leq C
\end{equation}
for some $C>0$. Hence, Gronwall gives
\begin{equation}
\label{EqUBoundD}
||d^{\;\,\rho}_{ \beta}|_{\Phi((u, \theta),s)} || \leq C \qquad \textnormal{ uniformly for } (u, \theta) \in B_\rho\big((u_0, \theta_0)\big) \textnormal{ and } s \in [-\frac{1}{2}, 0) \;.
\end{equation}
Integrating the parallel transport equation we obtain for all $s_0 \leq s_1 < s_2 <0$
\begin{equation*}
d^{\;\,\rho}_{ \beta}|_{\Phi((u, \theta),s_2)} = d^{\;\,\rho}_{ \beta}|_{\Phi((u, \theta),s_1)} - \int_{s_1}^{s_2}( d^{\;\,\alpha}_{ \beta}\mathbb{A}^{\;\,\rho}_{ \alpha})|_{\Phi((u, \theta),s)} \, ds
\end{equation*}
and combining this with \eqref{EqIntA} and \eqref{EqUBoundD} yields easily that $d^{\rho}_{\; \beta}|_{\Phi((u, \theta),s)}$ extends continuously to $(u, \theta, s) \in B_\rho\big((u_0, \theta_0) \big) \times [- \frac{1}{2}, 0]$. Finally, since $\Phi$ extends as a homeomorphism to this set and since the frame field $e_\alpha$ is continuous on $\Phi\Big(B_\rho \big((u_0, \theta_0)\big) \times [s_0, \frac{1}{2}]\Big)$, so is $d_\beta = d^\alpha_{\; \beta} e_\alpha$. This proves the claim.

Finally note that $\partial_s \Phi^\mu\big((u,\theta),s) = L^\mu|_{\Phi\big((u,\theta),s\big)}$ is uniformly bounded on $(u,\theta, s) \in B_\rho\big((u_0, \theta_0)\big) \times [-\frac{1}{2}, 0)$ in the $x^\mu$ (or $(u, \uu, \theta)$) coordinates. This verifies the last assumption in Proposition 5.11 in \cite{Sbie24a} so that we can conclude Step 2.
\newline

\textbf{Step 3:} \emph{Using the Lipschitz extension we construct vector fields $\hat{X}_i$, satisfying the assumptions of the theorem, such that \eqref{EqThmR} is uniformly bounded -- which gives the contradiction.}

First recall that $\id|_{\tilde{W}_<}$ extending as a $C^1$ isometric diffeomorphism to $\id|_{\tilde{W}_\leq} : \tilde{W}_\leq \to \hat{W}_\leq$ means that one can extend the isometry $\id|_{\tilde{W}_<}$ as a $C^1$ diffeomorphism to some open subset of $\tilde{W}$ containing $\tilde{W}_\leq$. So after making $\tilde{W}$ and $\hat{W}$ smaller if necessary we can assume that $\id|_{\tilde{W}_<}$ extends as a $C^1$-diffeomorphism to $\id|_{\tilde{W}} : \tilde{W} \to \hat{W}$.

Let us denote the coordinates induced by $\hat{\varphi} : \hat{U} \to (-\varepsilon_2, \varepsilon_2) \times (-\varepsilon_3, \varepsilon_3)^3$ by $y^\mu$. After making $\tilde{W}$ and $\hat{W}$ smaller if necessary we may assume that 
\begin{equation}
\label{EqBoundYCoord}
|\hat{g}_{\mu \nu} - \eta_{\mu \nu}| \leq \frac{1}{10} \qquad \textnormal{ and } \qquad | \rd_\kappa \hat{g}_{\mu \nu}| \leq \Lambda \quad \textnormal{ on } \hat{W}_<
\end{equation}
for some $0<\Lambda < \infty$.
As a consequence, if $\hat{Y}, \hat{Z}$ are smooth vector fields on $\hat{W}$ such that their components and the partial derivatives of those components with respect to the $y^\mu$ coordinates are uniformly bounded, then the components
\begin{equation}
\label{EqBoundChristoffelHat}
(\hat{\nabla}_{\hat{Y}}\hat{Z})^\mu = \hat{Y}^\kappa \frac{\rd}{\rd y^\kappa} \hat{Z}^\mu + \hat{\Gamma}^\mu_{\kappa \rho} \hat{Y}^\kappa \hat{Z}^\rho \leq C
\end{equation}
are uniformly bounded on $\hat{W}_<$.

We now make contact with assumption \eqref{EqThmR} of the theorem. Recall that by construction of the extension $\tilde{\iota} : M \hookrightarrow \tilde{M}$ we have $\overline{M} \subseteq \tilde{M}$ and $\tilde{p}=: \overline{p} \in \rd \overline{M}$. Moreover, $\overline{W} := \tilde{W} \cap \overline{M} = \tilde{W}_\leq$ is an open neighbourhood of $\overline{p}$ in $\overline{M}$. Thus, by the assumptions of the theorem there now exists $\overline{q} \in \overline{W} \cap \{\uu = 0\}$ and a compact neighbourhood $\overline{V} \subseteq \overline{W} = \tilde{W}_\leq$ of $\overline{q}$ such that for all $\uu' \leq 0$ close enough to $0$ the set $\overline{V} \cap \{\uu \leq \uu'\}$ is a smooth manifold with corners, and, moreover, there exist continuous vector fields $\overline{X}_i$ on $\overline{V}$, $i = 1, \ldots, 4$, and an $\varepsilon_0>0$ such that for any continuous vector fields $\hat{X}_i$ on $\overline{V}$ with $\sup_{\overline{V}} || \hat{X}_i - \overline{X}_i||_h < \varepsilon_0$ we have
$\big|\int\limits_{\overline{V}\cap \{\uu \leq \uu_k\}} R(\hat{X}_1, \hat{X}_2, \hat{X}_3, \hat{X}_4) \, \vol \big| \to \infty$
along a sequence $\uu_k \to 0$.
\begin{figure}[h] 
  \centering
  \def\svgwidth{11cm}
   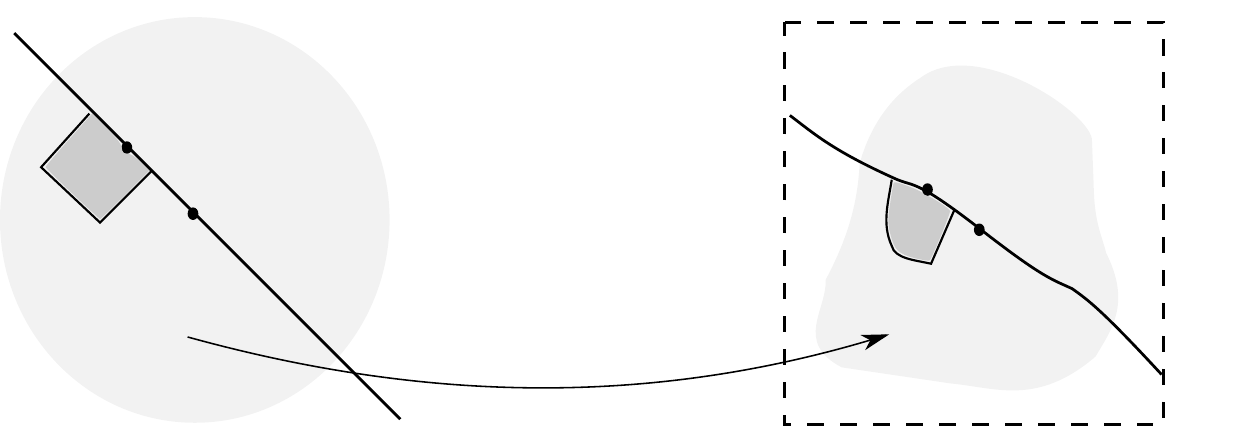 
      \caption{The $C^1$-identification of $\tilde{W}_\leq$ with $\hat{W}_\leq$.} \label{FigStep3}
\end{figure}

Using the Tietze extension theorem in coordinates, we first extend the vector fields $\overline{X}_i$ to continuous vector fields on all of $\tilde{W}$ and denote them by $\tilde{X}_i$. We then consider the vector fields $(\id|_{\tilde{W}})_* \tilde{X}_i$ on $\hat{W}$, which are continuous as well, since $(\id|_{\tilde{W}})_*$ is $C^1$. Furthermore, we extend the auxiliary Riemannian metric $h$ to $\tilde{M}$ by the same prescription as in \eqref{EqDefH} and call it $\tilde{h}$. Also $(\id|_{\tilde{W}})_*\tilde{h}$ is a continuous Riemannian metric on $\hat{W}$. Smoothing out $(\id|_{\tilde{W}})_*\tilde{X}_i$ on $\hat{W}$ with respect to the smooth structure of $\hat{M}$ (note that we have shown in Step 2 that the $C^1$-structures on $\tilde{W}$ and $\hat{W}$ are equivalent, but the $C^2$-structures are in general inequivalent) we can find smooth vector fields $\hat{Y}_i$ on $\hat{W}$ with 
$$\sup_{\hat{V}} || \hat{Y}_i - \id_* \tilde{X}_i||_{(\id|_{\tilde{W}})_*\tilde{h}} < \varepsilon_0 \;,$$
where $\hat{V} := \id|_{\tilde{W}}(\overline{V})$.
We now pull back the vector fields $\hat{Y}_i$ to $\tilde{W}$ to obtain continuous vector fields $\hat{X}_i := (\id|_{\tilde{W}}^{-1})_* \hat{Y}_i$ on $\tilde{W}$ satisfying  
\begin{equation} \label{EqModi}
\sup_{\overline{V}} || \hat{X}_i - \overline{X}_i||_h < \varepsilon_0 \;.
\end{equation}
Thus, along a sequence $\uu_k \to 0$ we have by assumption \eqref{EqThmR}
\begin{equation}
\label{EqRBlow}
\underbrace{\Big{|} \int\limits_{\overline{V} \cap \{\uu < \uu_k\}} R(\hat{X}_1, \hat{X}_2, \hat{X}_3, \hat{X}_4) \,\vol \Big{|}}_{=:I_k} \to \infty  \;.
\end{equation}
On the other hand we estimate (see justification below)
\begin{align}
I_k &= \Big{|} \int\limits_{\id(\overline{V} \cap \{\uu < \uu_k\})} \hat{R}(\hat{Y}_1, \hat{Y}_2, \hat{Y}_3, \hat{Y}_4) \,\hat{\vol} \Big| \notag \\
&= \Big| \int\limits_{\id(\overline{V} \cap \{\uu < \uu_k\})} \hat{g}\Big( \hat{\nabla}_{\hat{Y}_1}(\hat{\nabla}_{\hat{Y}_2} \hat{Y}_3) - \hat{\nabla}_{\hat{Y}_2}(\hat{\nabla}_{\hat{Y}_1}\hat{Y}_3) - \hat{\nabla}_{[\hat{Y}_1, \hat{Y}_2]} \hat{Y}_3, \hat{Y}_4\Big) \, \hat{\vol} \Big| \notag \\
&\leq C + \Big| \int\limits_{\id(\overline{V} \cap \{\uu < \uu_k\})} \Big[ \hat{Y}_1 \big(\hat{g}(\hat{\nabla}_{\hat{Y}_2}\hat{Y}_3,\hat{Y}_4) \big) - \hat{g}(\hat{\nabla}_{\hat{Y}_2} \hat{Y}_3, \hat{\nabla}_{\hat{Y}_1}\hat{Y}_4) - \hat{Y}_2 \big(\hat{g}(\hat{\nabla}_{\hat{Y}_1} \hat{Y}_3, \hat{Y}_4) \big) + \hat{g} ( \hat{\nabla}_{\hat{Y}_1} \hat{Y}_3, \hat{\nabla}_{\hat{Y}_2}\hat{Y}_4) \Big] \, \hat{\vol} \Big| \notag \\
&\leq C + \Big| \int\limits_{\id(\overline{V} \cap \{\uu < \uu_k\})} d \Big( \hat{g}(\hat{\nabla}_{\hat{Y}_2} \hat{Y}_3, \hat{Y}_4) \cdot \hat{Y}_1 \lrcorner \hat{\vol} \Big) - \hat{g} (\hat{\nabla}_{\hat{Y}_2} \hat{Y}_3, \hat{Y}_4) \cdot \mathrm{div} \hat{Y}_1 \cdot \hat{\vol} \Big|  \notag \\
&\qquad + \Big| \int\limits_{\id(\overline{V} \cap \{\uu < \uu_k\})} d \Big( \hat{g}(\hat{\nabla}_{\hat{Y}_1} \hat{Y}_3, \hat{Y}_4) \cdot \hat{Y}_2 \lrcorner \hat{\vol} \Big) - \hat{g} (\hat{\nabla}_{\hat{Y}_1} \hat{Y}_3, \hat{Y}_4) \cdot \mathrm{div} \hat{Y}_2 \cdot \hat{\vol} \Big| \notag \\
&\leq C +\Big| \int\limits_{\rd \id (\overline{V} \cap \{\uu \leq \uu_k\})} \hat{g}(\hat{\nabla}_{\hat{Y}_2} \hat{Y}_3, \hat{Y}_4) \cdot \hat{Y}_1 \lrcorner \hat{\vol} \Big| +\Big| \int\limits_{\rd \id (\overline{V} \cap \{\uu \leq \uu_k\})} \hat{g}(\hat{\nabla}_{\hat{Y}_1} \hat{Y}_3, \hat{Y}_4) \cdot \hat{Y}_2 \lrcorner \hat{\vol} \Big| \label{EqForRem} \\ 
&\leq C + C \sum_{i= 1,2} \Big( \int\limits_{\id(\overline{V} \cap \{\uu = \uu_k\})} \big| \hat{Y}_i \lrcorner \hat{\vol} \big| + \int\limits_{\rd \id(\overline{V})} \big|\hat{Y}_i \lrcorner \hat{\vol}\big| \Big) \notag \\
&\leq C \;. \notag
\end{align}
Here, $\hat{\vol}$ denotes the Lorentzian volume form on $\hat{U}$ with respect to $\hat{g}$ such that $(\id|_{\tilde{W}_<})^* \hat{\vol} = \vol$.
Relying on the smoothness of the vector fields $\hat{Y}_i$ we have used the definition of the curvature tensor in the second step. In the third step we used the product rule on the first two terms and bounded the last term using the compactness of $\hat{V}$, the smoothness of $[\hat{Y}_1, \hat{Y}_2]$ on $\hat{W}$, and \eqref{EqBoundYCoord} and \eqref{EqBoundChristoffelHat}. In the fourth step the first order terms were bounded as before. For the second order terms we first used, for a smooth function $f$, the relation $\hat{Y}_i(f) \cdot \hat{\vol} = \mathcal{L}_{\hat{Y}_i} (f \cdot \hat{\vol}) - f \mathcal{L}_{\hat{Y}_i} \hat{\vol} = \mathcal{L}_{\hat{Y}_i} (f \cdot \hat{\vol}) - f \mathrm{div} \hat{Y}_i \cdot \hat{\vol}$ and then Cartan's formula, where $\hat{Y}_i \lrcorner \hat{\vol}(\cdot, \cdot, \cdot) := \hat{\vol}(\hat{Y}_i, \cdot, \cdot, \cdot)$. In the fifth step we use Stokes theorem for manifolds with corners and bound again the first order terms as before. In the sixth step we bound again the first order terms as before, estimate  the integral of the three-form over the boundary by the corresponding density, and also split the boundary integral into a $k$-dependent part and estimate the remaining part by the integral over all of $\rd \id(\overline{V})$. Finally, the last integral is an integral of a continuous $3$-density over a compact piecewise $C^1$ boundary and thus finite, while to see the uniform boundedness of the first integral we may pull it back via $\id$ to obtain 
$$\int\limits_{\overline{V} \cap \{\uu = \uu_k\}} | \hat{X}_i \lrcorner \tilde{\vol}| = \int\limits_{\overline{V} \cap \{\uu = \uu_k\}}  \sqrt{- \mathrm{det} \tilde{g}} |\hat{X}_i^{\uu}| \, du d\theta^1d\theta^2 \;.$$ Since the integrand and the domain of integration are uniformly bounded in $k$, this integral is also uniformly bounded. This gives the contradiction to \eqref{EqRBlow} and thus concludes the proof of the main theorem.
\end{proof}

\begin{proof}[Proof of Remark \ref{RemVariation}]
Step 1 and Step 2 are identical. In Step 3, the only modification is that instead of only extending and then smoothing out the four vector fields $\overline{X}_i$ we extend and smooth out the eight vector fields $\overline{X}_i^{(j)}$ to obtain in place of \eqref{EqModi}
\begin{equation*}
\sup_{\overline{V}} || \hat{X}_i^{(j)} - \overline{X}_i^{(j)}||_h < \varepsilon_0 \;.
\end{equation*}
Then, along a sequence $\uu_k \to 0$ we have by assumption \eqref{EqThmR2}
\begin{equation*}
\Big{|} \int\limits_{\overline{V} \cap \{\uu < \uu_k\}} R(\hat{X}_1^{(1)}, \hat{X}_2^{(1)}, \hat{X}_3^{(1)}, \hat{X}_4^{(1)}) + \mathrm{i}R(\hat{X}_1^{(2)}, \hat{X}_2^{(2)}, \hat{X}_3^{(2)}, \hat{X}_4^{(2)}) \,\vol \Big{|} \to \infty \;.
\end{equation*}
But this implies that, possibly along a subsequence, we must have
\begin{equation}\label{EqThmR2}
\Big{|} \int\limits_{\overline{V} \cap \{\uu < \uu_k\}} R(\hat{X}_1^{(j)}, \hat{X}_2^{(j)}, \hat{X}_3^{(j)}, \hat{X}_4^{(j)}) \,\vol \Big{|} \to \infty 
\end{equation}
for $j=1$ or $j=2$. This now again puts us into the setting of \eqref{EqRBlow} and the remainder of the proof proceeds as before.
\end{proof}

\begin{remark} \label{RemRate}
The method introduced in Step 3 of the proof also gives a possible approach for obtaining an inextendibility statement with $\tilde{g} \in C^0$ and $\partial \tilde{g} \in L^2_{\loc}$: if the extension is assumed to only have this regularity, then the boundary term in \eqref{EqForRem} is not necessarily bounded. With a minor modification, taking care of the boundary terms in the angular and the $u$ direction, one can now square the whole estimate up to this point and integrate again in $\underline{u}_k$ to get an expression which has to be finite for $\tilde{g}$ in the above regularity class. If on the other hand the corresponding doubly integrated (and squared) curvature expression is assumed infinite, one again obtains the desired contradiction. The main problem with implementing this idea of proof is that the analogue of Step 2 is not known to hold; below Lipschitz regularity the methods of \cite{Sbie24a} break down. Indeed, in \cite{CaSbi24} it is shown that weak null singularities as considered here admit continuous extensions for which the identification map $\id$ does not extend as a $C^1$-regular diffeomorphism to the boundary!
\end{remark}

\section{Examples and counterexamples} \label{SecEx}

As discussed in the introduction, the main application of Theorem \ref{MainThm} is  to the $C^{0,1}_{\loc}$-inextendibility of weak null singularities occurring in the interior of perturbations of  rotating black holes. Since the Kerr family of black holes is expected to comprise all stationary asymptotically flat vacuum black holes, such weak null singularities are expected to occur in the interior of any generic asymptotically flat vacuum black hole. The gauge, with respect to which Theorem \ref{MainThm} is formulated, is that of \cite{DafLuk17} -- so that the inextendibility theorem is directly applicable in this context once the integrated curvature blow-up condition \eqref{EqThmR} (and \eqref{EqAssumpB}) is established. Other (prospective) applications are to weak null singularities without symmetries for the Einstein-Euler equations \cite{Song25}, spherically symmetric weak null singularities for the Einstein-Maxwell-(charged) scalar field/null dust system (see discussion below Example \ref{ExToyR}), and Milne-like spacetimes for a specific range of scale factor, cf.\ \cite[Theorem 3.4]{Ling20}.

In this section we give an explicit simple toy-example in spherical symmetry to which Theorem \ref{MainThm} is applicable.\footnote{Recall that the advantage of the method presented in this paper is that it is easy to handle even in spacetimes that \emph{do not} have any symmetry. So the main purpose of the spherically symmetric examples presented in this section is to illustrate the new \emph{method} in itself in its simplest setting.}  A modification of this toy-example also shows that the condition \eqref{EqThmR} on curvature blow-up is optimal in the sense that one can neither bring the absolute value inside the integral nor can one demand \eqref{EqThmR} for the $\overline{X}_i$ only instead of the perturbed vector fields $\tilde{X}_i$ -- both cases are compatible with the existence of a $C^{0,1}_{\loc}$-extension. This proves Remark \ref{RemStructureRCond}. 

\begin{example} \label{ExToy}
Consider \eqref{EqGDN} with $\Omega \equiv 1$, $b \equiv 0$, and $\gamma = r^2(\uu)(d\theta^2 + \sin^2 \theta d \varphi^2) = r^2(\uu) \mathring{\gamma}$, where $(\theta, \varphi)$ are the standard coordinates on $\mathbb{S}^2$, $\mathring{\gamma}$ is the round metric on $\Sp^2$, and $r : [-1,0) \to (0, \infty)$ is a smooth function which will be chosen in different ways in the next examples. Thus, in $(u, \uu, \theta, \varphi)$ coordinates the metric \eqref{EqGDN} and the inverse metric take the form
\begin{equation*}
g = \begin{pmatrix}
0 & -2 & 0 & 0 \\ -2 & 0 & 0 & 0 \\ 0 & 0 & r^2(\uu) & 0 \\ 0 & 0 & 0 & r^2(\uu) \sin^2 \theta
\end{pmatrix}, \qquad
g^{-1} = \begin{pmatrix}
0 & - \frac{1}{2} & 0 & 0 \\ - \frac{1}{2} & 0 & 0 & 0 \\ 0 & 0 & \frac{1}{r^2(\uu)} & 0 \\ 0 & 0 & 0 & \frac{1}{r^2(\uu) \sin^2 \theta} 
\end{pmatrix}\;,
\end{equation*}
the non-vanishing components of the Christoffel symbols, modulo those obtainable from symmetry, are computed to read
\begin{equation*}
\Gamma^u_{AB} = \frac{1}{2} r \rd_{\uu} r \mathring{\gamma}_{AB} \;, \qquad  \Gamma^A_{B\uu} = \frac{1}{r} \rd_{\uu} r \delta^A_{\; \,B}\;, \qquad \Gamma^\theta_{\varphi \varphi} = - \sin \theta \cos \theta\;, \qquad \Gamma^\varphi_{\varphi \theta} = \cot \theta \;,
\end{equation*}
and the non-vanishing curvature components, modulo those obtainable from symmetry, are given by
\begin{equation}\label{EqToyRExp}
R_{\uu \theta \theta \uu} = r \rd_{\uu}^2 r\;, \qquad R_{\uu \varphi \varphi \uu} =  \sin^2 \theta \cdot r \rd_{\uu}^2 r  \;, \qquad R_{\theta \varphi \varphi \theta} = -r^2 \sin^2 \theta \;.
\end{equation}
Furthermore, we have $e_3 = \rd_u$ and $e_4 = \rd_{\uu}$ and
\begin{equation}\label{EqRicciCoefToy}
\chi_{AB} = r \rd_{\uu} r \cdot \mathring{\gamma}_{AB}\;, \qquad \underline{\chi}_{AB} = 0\;, \qquad \eta_A = \underline{\eta}_A = 0\;, \qquad \underline{\omega}=0 \;.
\end{equation}
Finally, recalling the definition of the norm $|\chi|_\gamma$ below \eqref{EqZeta}, we compute 
\begin{equation}
\label{EqNormChiToy}
|\chi|_{\gamma} = \frac{\sqrt{2}}{r} |\rd_{\uu} r|\;.
\end{equation}
\end{example}
We now give a very simple toy-example of a spacetime with a weak null singularity and illustrate the application of Theorem \ref{MainThm}.
\begin{example} \label{ExToyR}
In the setting of Example \ref{ExToy} we set $r(\uu) := 1 + \frac{1}{\big|\log (|\uu|)-1\big|^{p_0-1}}$ for some $p_0 >1$. Note that $r(\uu)$ extends continuously to $\uu =0$ and hence $g$ does as well. Thus, the first assumption of Proposition \ref{PropGeod} is satisfied. That the second assumption \eqref{EqPropAssumRicci} is satisfied follows from \eqref{EqRicciCoefToy}, \eqref{EqNormChiToy} together with $r(\uu) \geq 1$ and $\rd_{\uu} r = \frac{p_0-1}{\uu \cdot \big| \log(|\uu|)-1 \big|^{p_0}} <0$. Assumption \eqref{EqAssumpB} in Theorem \ref{MainThm} is trivially satisfied. 

To apply Theorem \ref{MainThm}, let now $\overline{p} \in \rd \overline{M}$ and an open neighbourhood $\overline{W} \subseteq \overline{M}$ of $\overline{p}$ be given. After a rotation of the spherical coordinate system $(\theta, \varphi)$ we can without loss of generality assume that $\overline{p} = (u_{\overline{p}}, 0, \frac{\pi}{2}, \pi)$ with $u_{\overline{p}} \in (-1,1)$. We choose $\overline{q} := \overline{p}$ and set
$$\overline{V}:= [u_{\overline{p}} - \delta, u_{\overline{p}} + \delta] \times [- \delta, 0] \times [ \frac{\pi}{2} - \delta, \frac{\pi}{2} + \delta] \times [ \pi - \delta, \pi + \delta] \;.$$
For $\delta>0$ sufficiently small we have $\overline{V} \subseteq \overline{W}$. Clearly, $\overline{V} \cap \{\uu \leq \uu'\}$ is a compact smooth manifold with corners for $\uu' \leq 0$ close enough to $0$. Now, the vector fields $\overline{X}_1 = \overline{X}_4 = \rd_{\uu}$ and $\overline{X}_2 = \overline{X}_3 = \rd_\theta$ are continuous on $\overline{V}$. We will determine $0<\varepsilon_0 <1$ below -- for now let $\hat{X}_i$, $i \in \{1,2,3,4\}$, be continuous vector fields on $\overline{V}$ with $\sup_{\overline{V}}|| \hat{X}_i - \overline{X}_i||_h < \varepsilon_0$. By \eqref{EqToyRExp},  we then have
\begin{equation} \label{EqEvalR}
R(\hat{X}_1, \hat{X}_2, \hat{X}_3, \hat{X}_4) = \underbrace{R(\overline{X}_1, \overline{X}_2, \overline{X}_3, \overline{X}_4)}_{=r \rd_{\uu}^2 r} + \mathcal{O}(\varepsilon_0) \cdot r \rd_{\uu}^2 r + \mathcal{O}(\varepsilon_0) \cdot \sin^2 \theta \cdot r \rd_{\uu}^2 r + \mathcal{O}(\varepsilon_0) \cdot r^2 \sin^2 \theta\;,
\end{equation}
where $\mathcal{O}(\varepsilon_0)$ stands for a function which, in absolute value, is bounded by $C \cdot \varepsilon_0$, where $C>0$ is a numerical constant independent of $\varepsilon_0$.

We now compute $$\rd_{\uu}^2 r(\uu) =  - \frac{p_0 -1}{\uu^2 \big| \log(|\uu|) -1 \big|^{p_0}} + \frac{(p_0 -1) p_0}{\uu^2 \big| \log (|\uu|) -1 \big|^{p_0 + 1}} \;.$$
Without loss of generality we can assume that $\delta>0$ is so small that for $- \delta \leq \uu < 0$ we have $$ - \frac{p_0 -1}{\uu^2 \big| \log(|\uu|) -1 \big|^{p_0}} \leq \rd_{\uu}^2 r(\uu) \leq - \frac{1}{2} \cdot \frac{p_0 -1}{\uu^2 \big| \log(|\uu|) -1\big|^{p_0}}\;.$$ Moreover, by our definition of $r$ we clearly have $1 \leq r \leq 2$. It thus follows from \eqref{EqEvalR} that we can choose $0 < \varepsilon_0 <1$ so small that 
$$R(\hat{X}_1, \hat{X}_2, \hat{X}_3, \hat{X}_4) \leq  - \frac{1}{4} \cdot \frac{p_0 -1}{\uu^2 \big| \log(|\uu|) -1\big|^{p_0}}\;.$$
This bound now implies \eqref{EqThmR} for any sequence $\uu_k \to 0$. Thus, the conclusion of Theorem \ref{MainThm} holds, i.e., there are no $C^{0,1}_{\loc}$-extensions of $(M,g)$ `across $\{\uu = 0\}$'.
\end{example}
Of course, the spacetime in Example \ref{ExToyR} does not arise from the Einstein equations coupled to a suitable matter model in a physically interesting situation. However, slightly more complicated spherically symmetric weak null singularities -- with $\Omega(u, \uu)$ and $r(u, \uu)$ -- do arise in the study of the spherically symmetric Einstein equations coupled to suitable matter models, see for example \cite{His81}, \cite{PoiIs89}, \cite{Ori91}, \cite{Daf05a}, \cite{LukOh19I}, \cite{VdM18}. We refrain from computing here the curvature tensors and from showing how Theorem \ref{MainThm} applies in those settings, since a satisfactory $C^{0,1}_{\loc}$-inextendibility criterion tailored to those situations, with input only at the level of first derivatives of the metric, has already been presented in \cite{Sbie22a}.

We now continue to use Example \ref{ExToy} to discuss in which sense condition \eqref{EqThmR} is sharp. We start by showing that bringing the absolute value in \eqref{EqThmR} inside of the integral is compatible with the existence of a $C^{0,1}_{\loc}$-extension.
\begin{example} \label{ExCounterAbs}
In the setting of Example \ref{ExToy} we define $\rd_{\uu}^2r(\uu) := \frac{1}{\uu} \sin( \frac{1}{\uu})$ and fix $\rd_{\uu} r(\uu) := \int_{-1}^{\uu} \frac{1}{\uu'} \sin (\frac{1}{\uu'}) \, d\uu'$. We claim that $\rd_{\uu} r(\uu)$ is uniformly bounded in absolute value. To see this we use the substitution $x = - \frac{1}{\uu}$ to obtain
\begin{equation*}
\begin{split}
\int_{-1}^{\uu} \frac{1}{\uu'} \sin (\frac{1}{\uu'}) \, d\uu' &= \int_1^{x(\uu)} \frac{1}{x'} \sin(x') \, dx' = - \int_{1}^{x(\uu)} \frac{1}{x'} \frac{d}{dx'} \cos(x') \, dx' \\
 &= - \frac{1}{x(\uu)} \cos(x(\uu)) + 1 \cos(1) - \int_1^{x(\uu)} \frac{1}{(x')^2} \cos (x')\, dx' \;.
\end{split}
\end{equation*}
From this we obtain the crude bound $|\rd_{\uu} r | \leq 4$. We now define $r(\uu)= 5 + \int_{-1}^{\uu} \rd_{\uu} r(\uu') \, d\uu'$, which extends continuously to $\uu = 0$. Clearly, we have $r \geq 1$. In the $(u, \uu, \theta, \varphi)$-differentiable structure we can now even construct a Lipschitz extension of $(M,g)$ across $\{\uu = 0\}$ by simply setting $r(\uu) = r(0)$ for $\uu >0$.

On the other hand, using the vector fields $\overline{X}_1 = \overline{X}_4 = \rd_{\uu}$ and $\overline{X}_2 = \overline{X}_3 = \rd_\theta$ we obtain as in \eqref{EqEvalR}
$$|R(\hat{X}_1, \hat{X}_2, \hat{X}_3, \hat{X}_4) | \geq |r \rd_{\uu}^2 r| \cdot |1 + \mathcal{O}(\varepsilon_0)| - \mathcal{O}(\varepsilon_0) \;.$$
Choose now $\varepsilon_0>0$ such that $1 + \mathcal{O}(\varepsilon_0) \geq \frac{1}{2}$. Also use $r \geq 1$ and note that $$\int_{-1}^{\uu} |\rd_{\uu}^2 r|(\uu') \,d\uu' = \int_{-1}^{\uu} |\frac{1}{\uu'} \sin (\frac{1}{\uu'})| \, d\uu' = \int_1^{x(\uu)} |\frac{1}{x'} \sin(x')| \, dx'$$ is divergent.
\end{example}
The next example shows that it is essential to have the blow up \eqref{EqThmR} not just for the vector fields $\overline{X}_i$, but indeed also for vector fields $\hat{X}_i$ which are $C^0$-close to the $\overline{X}_i$ with respect to the metric $h$.
\begin{example} \label{ExCounterBar}
Again, we consider the setting of Example \ref{ExToy} and define $r(\uu) := 2 - \int_{-1}^{\uu} \cos (- \frac{1}{\uu'}) \, d\uu'$. Thus, we have $r \geq 1$ and $r(\uu)$ extends continuously to $\uu =0$. Moreover, we have $\rd_{\uu}r = - \cos(- \frac{1}{\uu})$ and hence we can construct again a $C^{0,1}_{\loc}$-extension across $\{\uu = 0\}$ by setting $r(\uu) = r(0)$ for $\uu > 0$.

Now, we compute $\rd_{\uu}^2 r (\uu) = \frac{1}{\uu^2} \sin (- \frac{1}{\uu})$ and we consider the vector fields $\overline{X}_2 = \overline{X}_3 = \frac{1}{\sqrt{r}} \rd_{\theta}$ and $\overline{X}_1 = \overline{X}_4 = \sqrt{1 - \uu \sin(\frac{1}{\uu})} \rd_{\uu}$, which are continuous on $\overline{M}$. Recalling $R_{\uu \theta \theta \uu} = r \rd_{\uu}^2 r$, we have
\begin{equation} \label{EqRSecCount}
R(\overline{X}_1, \overline{X}_2, \overline{X}_3, \overline{X}_4) = \rd_{\uu}^2 r(\uu) \cdot  \big(1 - \uu \sin(\frac{1}{\uu})\big) = \sin(- \frac{1}{\uu})\big(1 - \uu \sin (\frac{1}{\uu}) \big) \frac{1}{\uu^2} \;.
\end{equation}
Using again the substitution $x = - \frac{1}{\uu}$, we note that the integral of \eqref{EqRSecCount} in $\uu$ is unbounded:
\begin{equation*}
\begin{split}
\int_{-1}^{\uu} \sin(- \frac{1}{\uu'})\big(1 - \uu' \sin (\frac{1}{\uu'})& \big) \frac{1}{(\uu')^2} \, d\uu' = \int_1^{x(\uu)} \sin(x') \big(1 - \frac{1}{x'} \sin(x') \, dx'\\
 &= - \cos(x(\uu)) + \cos(1) - \int_{1}^{x(\uu)} \frac{1}{x'} \sin^2(x') \, dx' \to - \infty \qquad \textnormal{for } \uu \to 0\;.
\end{split}
\end{equation*}
This entails that condition \eqref{EqThmR} is satisfied for the \emph{particular choice} of vector fields $\overline{X}_i$, $i = 1,2,3,4$, despite the existence of a Lipschitz extension. 

Note, however, that this divergence is unstable under arbitrarily small $C^0$-perturbations of the vector fields $\overline{X}_i$: consider the continuous vector fields $\hat{X}_1 = \hat{X}_4 = \rd_{\uu}$ and $\hat{X}_2 = \hat{X}_3 = \overline{X}_2 = \overline{X}_3$. Note that we have $|| \hat{X}_1 - \overline{X}_1||_h \to 0$ for $\uu \to 0$ and thus, by suitably modifying these vector fields for $\uu \leq -\delta$, where  $\delta >0$, the perturbed vector fields can be arranged to be arbitrarily close to the unperturbed ones. Moreover, we have
$$\int_{- \delta}^{\uu}  R(\hat{X}_1, \hat{X}_2, \hat{X}_3, \hat{X}_4) \, d\uu' = \int_{- \delta}^{\uu} \sin(- \frac{1}{\uu'}) \frac{1}{(\uu')^2} \, d\uu' = \int_{\frac{1}{\delta}}^{x(\uu)} \sin (x') \, dx' = - \cos(x(\uu)) + \cos( \frac{1}{\delta})\;,$$
which is uniformly bounded.
\end{example}

\bibliographystyle{acm}
\bibliography{Bibly}

\begin{thebibliography}{10}

\bibitem{BraSmith95}
{\sc Brady, P.~R., and Smith, J.~D.}
\newblock {Black Hole Singularities: A Numerical Approach}.
\newblock {\em Phys. Rev. Lett. 75\/} (Aug 1995), 1256--1259.

\bibitem{CaSbi24}
{\sc Cameron, P., and Sbierski, J.}
\newblock {On the uniqueness of continuous spacetime extensions in $1+1$
  dimensions with applications to weak null singularities}.
\newblock {\em Forthcoming\/} (2025).

\bibitem{ChrisKlainStability}
{\sc Christodoulou, D., and Klainerman, S.}
\newblock {\em {The Global Nonlinear Stability of the Minkowski Space}}.
\newblock Princeton University Press, 1993.

\bibitem{ChrusGra12}
{\sc Chru\'sciel, P., and Grant, J.}
\newblock {On Lorentzian causality with continuous metrics}.
\newblock {\em Class. Quantum Grav. 29\/} (2012).

\bibitem{ChrusKli18}
{\sc Chru\'sciel, P., and Klinger, P.}
\newblock {The annoying null boundaries}.
\newblock {\em J. Phys. Conf. Ser. 968\/} (2018).

\bibitem{Daf05a}
{\sc Dafermos, M.}
\newblock {The interior of charged black holes and the problem of uniqueness in
  general relativity}.
\newblock {\em Comm. Pure Appl. Math. 58\/} (2005), 445--504.

\bibitem{DafLuk17}
{\sc Dafermos, M., and Luk, J.}
\newblock {The interior of dynamical vacuum black holes I: The $C^0$-stability
  of the Kerr Cauchy horizon}.
\newblock {\em arXiv:1710.01772\/} (2017).

\bibitem{GalLin16}
{\sc Galloway, G., and Ling, E.}
\newblock {Some Remarks on the $C^0$-(in)extendibility of Spacetimes}.
\newblock {\em Ann. Henri Poincar\'e 18}, 10 (2017), 3427--3477.

\bibitem{GalLinSbi17}
{\sc Galloway, G., Ling, E., and Sbierski, J.}
\newblock {Timelike completeness as an obstruction to $C^0$-extensions}.
\newblock {\em Comm. Math. Phys. 359}, 3 (2018), 937--949.

\bibitem{GrafLing18}
{\sc Graf, M., and Ling, E.}
\newblock {Maximizers in Lipschitz spacetimes are either timelike or null}.
\newblock {\em Class. Quantum Grav. 35}, 8 (2018).

\bibitem{GraVdBS24}
{\sc Graf, M., and van~den Beld-Serrano, M.}
\newblock {$C^0$-inextendibility of FLRW spacetimes within a subclass of
  axisymmetric spacetimes}.
\newblock {\em arXiv:2409.13799\/} (2024).

\bibitem{GraKuSa19}
{\sc Grant, J., Kunzinger, M., and S\"amann, C.}
\newblock {Inextendibility of spacetimes and Lorentzian length spaces}.
\newblock {\em Ann. Glob. Anal. Geom. 55\/} (2019), 133--147.

\bibitem{Gur24}
{\sc Gurriaran, S.}
\newblock {Precise asymptotics of the spin +2 Teukolsky field in the Kerr black
  hole interior}.
\newblock {\em arXiv:2409.0267\/} (2024).

\bibitem{HawkEllis}
{\sc Hawking, S., and Ellis, G.}
\newblock {\em The large scale structure of space-time}.
\newblock Cambridge University Press, 1973.

\bibitem{His81}
{\sc Hiscock, W.}
\newblock {Evolution of the interior of a charged black hole}.
\newblock {\em Phys. Rev. Lett. 83A\/} (1981), 110--112.

\bibitem{LeeSmooth}
{\sc Lee, J.}
\newblock {\em {Introduction to Smooth Manifolds}}.
\newblock Springer, 2002.

\bibitem{Ling20}
{\sc Ling, E.}
\newblock {The Big Bang is a Coordinate Singularity for $k = -1$ inflationary
  FLRW Spacetimes}.
\newblock {\em Foundations of Physics 50\/} (2020), 385--428.

\bibitem{Ling24}
{\sc Ling, E.}
\newblock {The $C^0$-inextendibility of some spatially flat FLRW spacetimes}.
\newblock {\em arXiv:2404.08257v1\/} (2024).

\bibitem{Luk18}
{\sc Luk, J.}
\newblock {Weak null singularities in general relativity}.
\newblock {\em Journal of the American mathematical society 31}, 1 (2018),
  1--63.

\bibitem{LukOh19I}
{\sc Luk, J., and Oh, S.-J.}
\newblock {Strong cosmic censorship in spherical symmetry for two-ended
  asymptotically flat initial data I. The interior of the black hole region}.
\newblock {\em Annals of Math. 190}, 1 (2019), 1--111.

\bibitem{McNam78}
{\sc McNamara, J.}
\newblock {Instability of black hole inner horizons}.
\newblock {\em Proc. R. Soc. Lond. A 358\/} (1978), 499--517.

\bibitem{Mie24}
{\sc Miethke, B.}
\newblock {$C^0$-inextendibility of the Kasner spacetime}.
\newblock {\em Master thesis, University of Vienna (also arXiv:2408.05257)\/}
  (2024).

\bibitem{MinSuhr19}
{\sc Minguzzi, E., and Suhr, S.}
\newblock {Some regularity results for Lorentz-Finsler spaces}.
\newblock {\em Ann. Glob. Anal. Geom. 56\/} (2019), 597--611.

\bibitem{Ori91}
{\sc Ori, A.}
\newblock {Inner Structure of a Charged Black Hole: An Exact Mass-Inflation
  Solution}.
\newblock {\em Phys. Rev. Lett. 67\/} (1991).

\bibitem{Ori00}
{\sc Ori, A.}
\newblock {Strength of curvature singularities}.
\newblock {\em Phys. Rev. D 61\/} (2000).

\bibitem{PoiIs89}
{\sc Poisson, E., and Israel, W.}
\newblock {Inner-Horizon Instability and Mass Inflation in Black Holes}.
\newblock {\em Phys. Rev. Lett. 63}, 16 (1989), 1663--1666.

\bibitem{Racz10}
{\sc R{\'a}cz, I.}
\newblock Spacetime extensions ii.
\newblock {\em Classical and Quantum Gravity 27}, 15 (2010), 155007.

\bibitem{Rein24}
{\sc Reintjes, M.}
\newblock {Strong Cosmic Censorship with bounded curvature}.
\newblock {\em Classical and Quantum Gravity 41}, 17 (jul 2024), 175002.

\bibitem{Sbie18}
{\sc Sbierski, J.}
\newblock {On the proof of the $C^0$-inextendibility of the Schwarzschild
  spacetime}.
\newblock {\em J. Phys. Conf. Ser. 968\/} (2018).

\bibitem{Sbie15}
{\sc Sbierski, J.}
\newblock {The $C^0$-inextendibility of the Schwarzschild spacetime and the
  spacelike diameter in Lorentzian geometry}.
\newblock {\em J. Diff. Geom. 108}, 2 (2018), 319--378.

\bibitem{Sbie22a}
{\sc Sbierski, J.}
\newblock {On holonomy singularities in general relativity and the
  ${C_{\mathrm{loc}}^{0,1}}$-inextendibility of space-times}.
\newblock {\em Duke Mathematical Journal 171}, 14 (2022), 2881 -- 2942.

\bibitem{Sbie23}
{\sc Sbierski, J.}
\newblock {Instability of the Kerr Cauchy Horizon Under Linearised
  Gravitational Perturbations}.
\newblock {\em Annals of PDE 9}, 7 (2023).

\bibitem{Sbie23b}
{\sc Sbierski, J.}
\newblock {The $C^0$-inextendibility of a class of FLRW spacetimes}.
\newblock {\em arXiv:2312.07443\/} (2023).

\bibitem{Sbie24a}
{\sc Sbierski, J.}
\newblock {Uniqueness and Non-Uniqueness Results for Spacetime Extensions}.
\newblock {\em International Mathematics Research Notices 2024}, 20 (09 2024),
  13221--13254.

\bibitem{Song25}
{\sc Song, Y.}
\newblock {Weak null singularity for the Einstein-Euler system}.
\newblock {\em arXiv:2506.16635v1\/} (2025).

\bibitem{VdM18}
{\sc Van~de Moortel, M.}
\newblock {Stability and Instability of the Sub-extremal
  Reissner--Nordstr{\"o}m Black Hole Interior for the
  Einstein--Maxwell--Klein--Gordon Equations in Spherical Symmetry}.
\newblock {\em Commun. Math. Phys. 360}, 1 (2018), 103--168.

\bibitem{VdM23}
{\sc Van~de Moortel, M.}
\newblock {The breakdown of weak null singularities inside black holes}.
\newblock {\em Duke Mathematical Journal 172}, 15 (2023), 2957 -- 3012.

\bibitem{VdM25}
{\sc Van~de Moortel, M.}
\newblock {The coexistence of null and spacelike singularities inside
  spherically symmetric black holes}.
\newblock {\em arXiv:2504.12370\/} (2025).

\end{thebibliography}

\end{document}